\documentclass{aa}
\usepackage{graphicx}
\usepackage{txfonts}
\usepackage{threeparttable}

\begin{document}

   \title{Magnetic cloud prediction model for forecasting space weather relevant properties of Earth-directed coronal mass ejections}

   \author{Sanchita Pal\inst{\ref{inst1}}\inst{\ref{inst2}}\and Dibyendu Nandy\inst{\ref{inst2}}\inst{\ref{inst3}}\and
Emilia~K.~J.~Kilpua\inst{\ref{inst1}}}

\institute{Department of Physics, University of Helsinki, P.O. Box 64, FI-00014 Helsinki, Finland\label{inst1} \and Center of Excellence in Space Sciences India, Indian Institute of Science Education and Research Kolkata, Mohanpur 741246, West Bengal, India\label{inst2} \and Department of Physical Sciences, Indian Institute of Science Education and Research Kolkata, Mohanpur 741246, West Bengal, India\\
\email{sanchita.pal@helsinki.fi}\label{inst3}
}

 \abstract
 {Coronal mass ejections (CMEs) are major eruptive events on the Sun that result in the ejection of large-scale magnetic clouds (MCs) in interplanetary space, consisting of plasma with enhanced magnetic fields whose direction changes coherently when measured in situ. The severity of CME-induced geomagnetic perturbations and space weather impacts depends on the direction and strength of the interplanetary magnetic field (IMF), as well as on the speed and duration of the passage of the magnetic cloud associated with the storm. The coupling between the heliospheric environment and Earth's magnetosphere is strongest when the IMF direction is persistently southward (i.e. negative $B_z$) for a prolonged period. Predicting the magnetic profile of such Earth-directed CMEs is therefore critical for estimating their space weather consequences; this remains an outstanding challenge, however. }
 {Our aim is to build upon and integrate diverse techniques towards the development of a comprehensive magnetic cloud prediction (MCP) model that can forecast the magnetic field vectors, Earth-impact time, speed, and duration of passage of solar storms.}
 {The configuration of a CME is approximated as a radially expanding force-free cylindrical structure. Combining near-Sun geometrical, magnetic, and kinematic properties of CMEs with the probabilistic drag-based model and cylindrical force-free model, we propose a method for predicting the Earth-arrival time, propagation speed, and magnetic vectors of MCs during their passage through 1 AU. Our model is able to predict the passage duration of the storm without recourse to computationally intensive time-dependent dynamical equations.}
 {Our method is validated by comparing the MCP model output with observations of ten MCs at 1 AU. In our sample, we find that eight MCs show a root mean square (rms) deviation smaller than 0.1 between the predicted and observed magnetic profiles, and the passage durations of seven MCs fall within the predicted range.}
 {Based on the success of this approach, we conclude that predicting the near-Earth properties of MCs based on an analysis and modelling of near-Sun CME observations is a viable endeavour with potential applications for the development of early-warning systems for space weather and enabling mitigation strategies.}

   \maketitle

\section{Introduction} \label{sec5.1}
Understanding space weather and its variability has become increasingly important as we rely more and more on satellites and space-reliant technologies such as communications and navigational networks. Space-weather-induced geomagnetic storms also impact interconnected electric power-grids, high-frequency communications and polar air-traffic, and they increase the drag on low-Earth orbiting satellites \citep[e.g.][] {schrijver2015understanding}. Ultimately, space weather originates in the creation of solar magnetic fields in the solar convection zone \citep{nandy2021a}, the surface emergence and evolution of solar active regions \citep{bhowmik2018}, and the subsequent evolution and dynamics of magnetic fields in the solar atmosphere and heliosphere over a range of timescales \citep{nandy2021b}. Solar flares and coronal mass ejections \citep[CMEs;][]{Webb2012}  are some of the most important drivers of severe space weather events. While approaches for forecasting solar flares are becoming increasingly feasible with the advent of machine-learning techniques \citep{sinha2022}, forecasting CMEs and their near-Earth manifestations remains highly challenging. CME flux rope (FR) structures can be observed in white-light coronagraphs near the Sun \citep[e.g.][]{Vourlidas2017} and in interplanetary space and near-Earth through in-situ observations   \citep[e.g.][]{Kilpua2017a}. When the twisted magnetic FR of a CME contains southward magnetic field components, magnetic reconnection with the Earth's magnetosphere ensues and leads to effective solar wind mass, momentum, and energy transfer  to the Earth's magnetosphere. This generates significant ring current enhancement and results in a geomagnetic storm \citep{tsurutani1988origin, gonzalez1999interplanetary}. Therefore, prior knowledge of the magnetic properties of Earth-directed CMEs is crucial for reliably predicting their geoeffectiveness and space weather impacts. Moreover, a long lead time in the forecasting of CME geoeffectiveness is desirable because it will allow space-reliant technologies and humanity  sufficient time to react to impending space weather hazards. 

  \par
  
The first step in estimating the geomagnetic response for a given CME is to estimate its initial parameters, including the structure, propagation direction, kinematics, and magnetic properties, soon after it originates on the Sun. Depending on the location of the CME source, CME properties are subject to projection effect \citep{JGRA:JGRA17179,howard2008kinematic}. By applying a forward- modelling technique \citep{2006ApJ...652..763T} to white-light CMEs observed simultaneously from different vantage points in space, the CME 3D morphology can be reproduced and deprojected geometrical parameters and kinematics can be estimated  \citep{2012SoPh..281..167B, 2013JGRA..118.6858S}. The next step is to evaluate how the initial parameters evolve after the CME is launched from the Sun. CMEs can experience changes in their space-weather-relevant properties and propagation direction \citep[e.g.][]{2017SSRv..212.1159M,Kilpua2019}. Firstly, CMEs expand during their interplanetary propagation \citep{1981JGR....86.6673B, burlaga1991intermittent}. \citet{demoulin2009causes} demonstrated that the rapid decrease in solar wind pressure with increasing distance from the Sun is the main driver of the radial expansion of CMEs.  The deflection of a CME can significantly change the latitude and longitude of its propagation direction \citep{2014SoPh..289.2141I,kay2015heliocentric}. In addition, CMEs can experience fast and strong rotation  \citep[e.g.][]{vourlidas2011first,2014SoPh..289.2141I}. These are common phenomena in the solar corona because strong magnetic forces exist in their  \citep{2014SoPh..289.2141I, 2015ApJ...805..168K}. Deflections and rotations also occur farther out in interplanetary space due to the interaction of a CME with the background solar wind magnetic fields \citep{ wang2004deflection} and preceding or following CMEs \citep{2004SoPh..222..329W, 2014JGRA..119.5117W}. Deflections can cause CMEs that were initially not heading towards the Earth to impact us or reroute Earth-directed CMEs away from our planet \citep[e.g.][]{Mostl2015}. The rotation in turn changes the magnetic field profile that finally impacts Earth, and thus influences the geoeffectivity \citep{palmerio2018coronal}. Interaction of a CME FR with the ambient open flux may also result in flux erosion impacting geoeffectiveness \citep{pal2020flux}. \citet{pal2022eruption} recently showed a large-scale magnetic structure that draped itself about a streamer-blowout CME flux rope at $\sim0.5$ AU, which  resulted in an erosion of almost $\sim20\%$ of the azimuthal magnetic flux of the CME.
All of these phenomena are a challenge for the prediction of near-Earth CME properties, which nonetheless is highly desirable.

\par
The distortion of the geometrical structure of a CME can be observed in coronagraphs. However, the influence of distortion of the CME magnetic structure is hard to estimate because the magnetic field cannot yet be reliably measured in the hot and tenuous corona. To estimate CME magnetic vectors, a few studies have used solar observations as input to 3D magnetohydrodynamic (MHD) models of CME evolution \citep{2004JGRA..109.2107M, shen2014evolution}. Although data-driven physical MHD models that can be used to predict the CME flux rope structure are desirable from the intellectual perspective, they are computationally expensive \citep{Manchester_2014} and do not have enough observations in the inner heliosphere to constrain their evolution.
\par
Various alternative semi-empirical modelling approaches have been proposed to predict the magnetic structure of CMEs. Using analytical and semi-analytical models that approximate CMEs as force-free cylindrical flux-ropes, several studies have predicted the magnetic structure of CME flux ropes as they arrive near the Earth's orbit \citep{2015SpWea..13..374S, kay2017using, mostl2018forward, sarkar2020observationally}. One key aspect of these models is that the magnetic properties of the CME flux rope are constrained as it leaves the Sun. The model by \citet{2015SpWea..13..374S} uses the  Bothmer-Schwenn scheme \citep{angeo-16-1-1998} as default. This scheme  relies on the hemispheric helicity rule \citep{pevtsov2003helicity}, which states that the northern (southern) hemisphere is dominated by magnetic structures with negative (positive) helicity sign and assumes that the orientation of the axial field of hte flux rope follows the polarity of the leading and trailing flux systems in active regions. The hemispheric rule applies only in a statistical sense, and intrinsic AR magnetic properties such as tilt orientation and twist themselves have a large scatter \citep{nandy2006magnetic}. Models of coronal field evolution and CME genesis based on active region properties often miss a large portion of the events \citep{yeates2010comparison}, implying a gap in our understanding. \citet{liu2014magnetic} showed that in only 60\% of the case is the hemispheric helicity rule followed in predicting the chirality of the CME flux rope. In addition, rising CME flux ropes interact with overlying coronal fields, and this interaction depends on the cycle phase \citep{cook2009solar}. This may alter the amount of magnetic flux and helicity as they reconnect with the overlying coronal arcades during the lift-off of the CME. This further compounds the prediction problem. \par
Some approaches have recently been proposed to make headway in the face of these challenges. The ForeCAT (Forecasting a CME’s Altered Trajectory) In situ Data Observer (FIDO) model developed by \citet{kay2017using} and the 3-Dimensional Coronal ROpe Ejection (3DCORE) model developed by \citet{mostl2018forward} take the expanding nature of the CME FR into account as it crosses the observing spacecraft in interplanetary space and use extreme-ultraviolet (EUV) observations to identify the FR foot point direction and chirality. The formulation of the Interplanetary Flux ROpe Simulator (INFROS) developed by \citet{sarkar2020observationally} includes the flux rope expansion in a way to remove expansion, the propagation speed, and the time of passage. It determines the time-varying axial field intensity and derives the axial field direction and FR chirality using EUV, H-alpha, and magnetogram observations of their sources. This approach allows constraining the FR parameters in a more realistic manner on an event-to-event basis \citep[see also][]{palmerio2017determining,Kilpua2019,PAL2021}. Most of the models discussed above do not incorporate a CME arrival time prediction, however, and forecast the flux rope passage time. \citet{2021ApJ...920...65P} recently evaluated the early evolution and forward-modelled the magnetic field of a slow stealth streamer-blowout CME up to a heliocentric distance of $\sim$0.5 AU by employing the modelling suite called Open Solar Physics Rapid Ensemble Information \citep[(OSPREI);][]{https://doi.org/10.1029/2021SW002914}. The suite couples three modules -- Forecasting a CME’s Altered Trajectory \cite[ForeCAT;][]{kay2013forecasting}, the Another Type of Ensemble Arrival Time Results \citep[ANTEATR;][]{2018JGRA..123.7220K}, and the ForeCAT In situ Data Observer \citep[FIDO][]{kay2017predicting}. While comparing the model results with the in-situ measurements at $\sim$0.5 AU, the study found encouraging agreement on arrival time, location of the spacecraft crossing, and magnetic profile.

\par
In this paper we present a comprehensive empirical modelling framework that builds upon our knowledge to predict various space-weather-relevant characteristics of a CME magnetic cloud near Earth. The model, which we call the CESSI-MCP model, does not involve the FR dimension, FR axial field intensity, FR arrival time, or speed as free parameters. It couples a few established models and techniques and uses the CME arrival information along with considerations of self-similar expansion to predict the magnetic profile and passage duration of the CME FR. In Section \ref{sec5.2} we describe the MCP model and outline the procedures and techniques for estimating the model inputs. In Section \ref{sec5.3} we validate our model using in-situ observed MC events. The results are critically assessed in Section \ref{sec5.4}, and we conclude with a summary and discussion in Section \ref{sec5.5}. 
 
\section{Method: Modelling MCs} \label{sec5.2}

\subsection{MCP model description}
 To examine the configuration of MCs, we assumed them to be force free \citep{1983NASCP.2280.731G}, that is, $\mathbf{J}=\alpha\mathbf{B}$, where $\mathbf{J}$ and $\mathbf{B}$ represent the current density and magnetic field vector, respectively. \citet{1986AdSpR...6..335M} first used this model, allowing $\alpha$ to vary with the distance from the MC centre to fit two MCs. Later,  \citet{burlaga1988magnetic} showed that $\alpha$ can be considered constant to describe a magnetic cloud to first order. For a constant $\alpha,$ the solutions of the force-free model in cylindrical coordinates were obtained by \citet{PhysRev.83.307}, where the axial ($B_{ax}$), azimuthal ($B_{az}$), and radial ($B_{rad}$) magnetic field components are given by\begin{equation}\label{eq5.1}
    B_{ax} = B_0 J_0 (\alpha \rho),
\end{equation}
\begin{equation}\label{eq5.2}
    B_{az} = H B_0 J_1 (\alpha \rho),
\end{equation} and
\begin{equation}\label{eq5.3}
 B_{\rho}= 0,
 \end{equation} 
 respectively. In Equation \ref{eq5.2}, $H$ represents the chirality of cylindrical FRs. The right- and left-handed chirality of FRs is indicated by $H=1$ and $H=-1$, respectively. The axial magnetic field intensity of FRs is represented by $B_0$. The zeroth- and first-order Bessel functions of the first kind are shown by $J_0$ and $J_1$, respectively. The parameter $\rho$ is the radial distance from MC axis, and $\alpha$ is related to the size of the FR. The value of $\alpha$ is chosen so that $\alpha R_{MC}= 2.41$, where 2.41 is the first zero of $J_0,$ and $R_{MC}$ is the radius of MC. 
 \par
 The field configuration described in Equations \ref{eq5.1} and \ref{eq5.2} is static. \citet{1981JGR..86..6673B} and \citet{burlaga1991intermittent} indicated that the expanding nature of MCs causes the smooth decrease in the solar wind speed and the low solar wind proton temperature during their intervals. \citet{demoulin2008expected} and \citet{demoulin2009causes} performed theoretical studies of the expansion of MCs. The studies concluded that MCs expand self-similarly, resulting in a linear radial velocity profile of MCs, and that the rate of the MC expansion is proportional to the MC radius. The expansion of MCs was first modelled by \citet{osherovich1993nonlinear}, followed by other studies, including \citet{marubashi1997interplanetary, hidalgo2003study, vandas2006comparison}, and \citet{2007AG..25..2453M}. These models are intended to fit the velocity magnitude profile of MCs. It is assumed that in an asymptotic limit, an FR expands radially with a speed
 \begin{equation} \label{eq5.4}
    V_{\rho}=\frac{\rho}{t+t_0},
\end{equation}
 where the force-free field configuration is maintained at any instant of time $t$ \citep{shimazu2002self, vandas2006comparison, vandas2015modeling}. In a self-similar expansion, the expansion time $t_0$ in Equation \ref{eq5.4} represents the time by which the expansion of FR has proceeded before it comes into contact with the spacecraft. If a self-similarly expanding MC changes its radius from its initial value $R_{MC}(0)$ to $R_{MC}(t)$ by the time $t$, the $R_{MC}(t)$ can be represented as
 $R_{MC}(t)= R_{MC}(0)(1+\frac{t}{t_0})$. Thus, for an expanding FR, $\alpha$ and $B_0$ become time dependent and are expressed as $\alpha= \frac{\alpha_0}{(1+\frac{t}{t_0})}$ and $B_0 = \frac{B'_0}{(1+\frac{t}{t_0})^2}$, where  $\alpha_0 = 2.41/R_{MC}(0)$ and $B'_0$ is the axial magnetic field intensity when the MC first encounters the spacecraft. Because of the expansion of the MC along the radial and axial directions, Equation \ref{eq5.1} and \ref{eq5.2} are modified as 
 \begin{equation} \label{eq5.5}
     B_{ax} =\frac{ B'_0 J_0 (\frac{\alpha_0}{(1+\frac{t}{t_0})} \rho)}{(1+\frac{t}{t_0})^2},
 \end{equation}
  \begin{equation} \label{eq5.6}
     B_{az} = H \frac{ B'_0 J_1 (\frac{\alpha_0}{(1+\frac{t}{t_0})} \rho)}{(1+\frac{t}{t_0})^2},
 \end{equation}
 where the force-free condition is assumed to be preserved throughout the propagation of MCs. 
 
\begin{figure}[]
 \centering
 \includegraphics[width=0.5\textwidth]{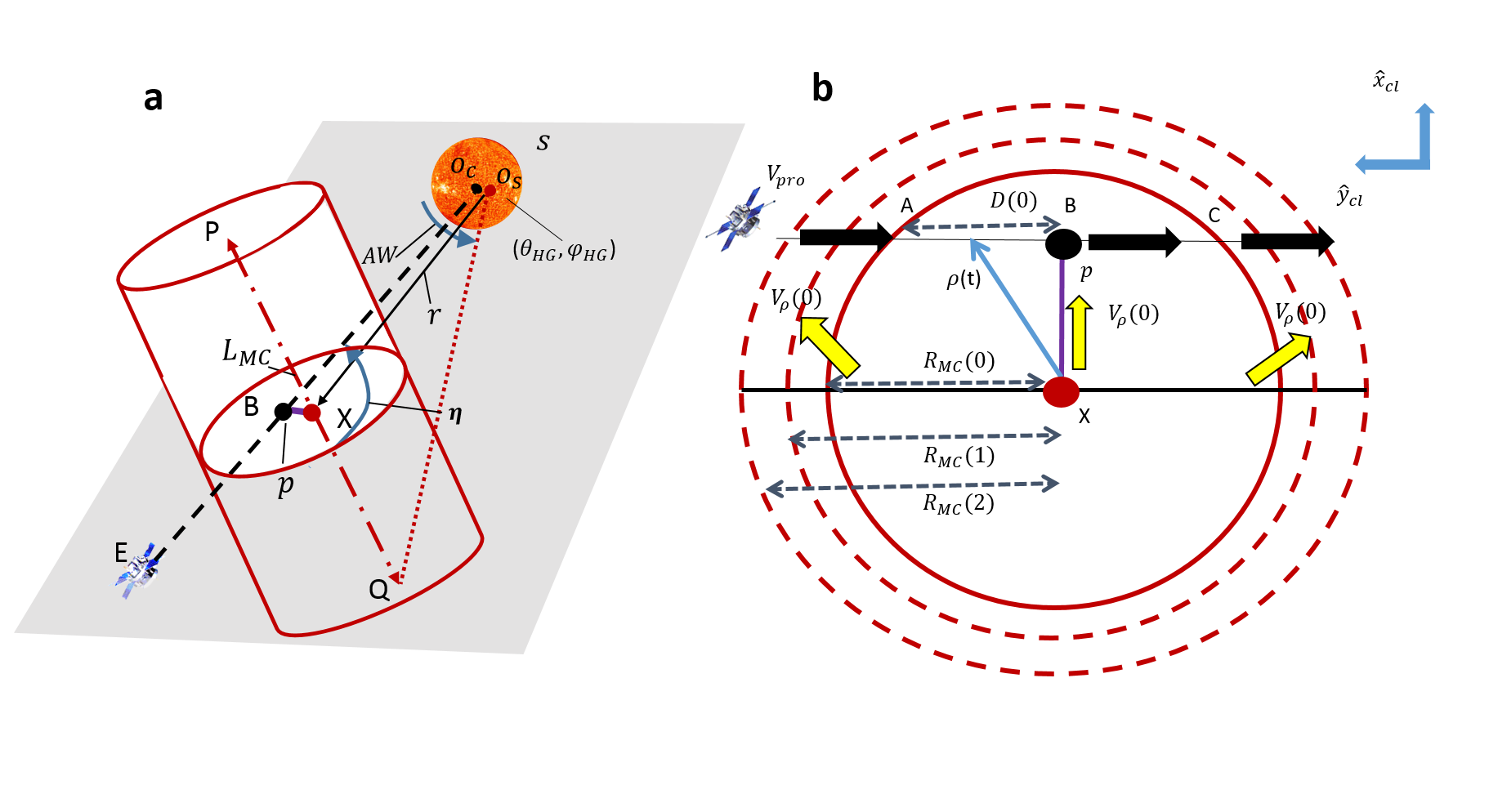}
 \caption{(a) Schematic of a cylindrical MC (solid red curve), the ecliptic plane (grey), and the position of the Sun S and the spacecraft E at L1. The $AW$, $\eta$, Sun-centre $O_c$ and the CME source location $O_s$ on the solar disk are labelled in the figure. The FR axis is shown as the dash-dotted red line. The Sun-Earth line $O_c$E is shown by the dashed black line and is on the ecliptic plane. The lines $O_s$X and PQ indicate the heliocentric distance $r$ and $L_{MC}$, respectively. The perpendicular distance $p$ is denoted by the violet line BX. The line BX is perpendicular to $O_c$E.  (b) The expanding circular cross-section of a cylindrical MC. X is the centre of the cross-section. At $t=0,$ the MC cross-sectional circumference is denoted by the red circle. The radial expansion of the MC with speed $V_{\rho}$ is in the direction of the yellow arrows, and in the FR frame of reference, the spacecraft propagates with a speed $V_{pro}$ towards the path indicated by black arrows. }.
\label{f5.1}
\end{figure}

 Knowledge of the perpendicular distance ($p$) between an MC axis and the location of a spacecraft performing the in-situ measurements of the MC is necessary for obtaining $\rho$. Figure \ref{f5.1} shows a cylindrical MC and its expanding cross-section. The MC expands with a radial velocity $V_{\rho}$ , and its axis propagates with a speed $V_{pro}$. In the FR frame of reference, it is assumed that the spacecraft propagates with the speed $V_{pro}$. In the in-bound and out-bound regions of the MC,  $V_{\rho}$ is added to and subtracted from the ambient solar wind speed to obtain $V_{pro}$ \citep{vandas2015modeling}. For $0< p <R_{MC}$ , the $\rho(t) = \sqrt{ p^2+(D^2(t)-V_{pro}(t)\times t)}$, where $D(t)= \sqrt{R^2_{MC}(t)-p^2}$. Thus at $t=0$, when an MC first encounters the spacecraft,  $\rho(0) = R_{MC}(0)$. Figure \ref{f5.1} is shown in the FR frame of reference where the spacecraft traverses the MC with a speed $V_{pro}$. The schematic shown in Figure \ref{f5.1}(a) represents the crossing of a spacecraft through an MC via a path indicated by the dashed black line. The MC axis is shown by the dash-dotted red line, and the circumference of the MC cross-section is indicated by a red circle. The heliocentric distance $r$ and the FR axis length $L_{MC}$ are indicated in the figure. The centres of the Sun and Earth are denoted by $O_c$ and E, respectively. The centre of the MC is indicated by X, and XB represents the distance $p$. In Figure \ref{f5.1}(b) the expanding MC cross-section is shown. At $t=0$, the solid red circle represents the MC cross-section circumference, where the distance AB is equivalent to $D(0)$. When the value of $V_{\rho}$ at $t=0$ is obtained, we can estimate the value of $t_0$ from $t_0= V^{t=0}_{\rho}/R_{MC}(0)$.  \par
 As mentioned in the Introduction, the reason for the CME expansion is mainly the decaying solar wind pressure that surrounds the CME. The expansion is rapid within a distance $\sim 0.4$ AU from the Sun and becomes moderate at a large distance \citep{2021A&A...649A..69S}. \citet{lepping2008estimates} formulated the scalar derivation of the expansion speed of FRs near the Earth that uses the FR width and propagation speed and the duration of the FR passage. The study analysed 53 MCs of standard profiles and obtained their expansion speed using different methods, namely the scalar method and vector determination. They found the most probable values of the expansion speed is about 30 km/s. \citet{nieves2018understanding} studied 337 ICMEs observed by the Wind spacecraft from 1995 to 2015 and reported that the MC expansion speed ranges from $\sim$ 56 to 271 km/s with a mean value of 28 km/s. In our study, we therefore estimated the expansion time $t_0$ by taking the initial radial expansion speed value $V^{t=0}_{\rho}$ as 28 km/s.    \par
 
 To obtain the model parameters, specifically $p$, $R_{MC}(0)$, $B'_0$ , and $H,$ the near-Sun observations of the associated CME FRs were used. As discussed in the Introduction, a significant deflection and rotation of CMEs regularly occur near the Sun, within $10 R_{\odot}$ \citep{kay2015heliocentric, lynch2009rotation}. We assumed here that the propagation direction, axis orientation, and chirality of CMEs obtained at a height $> 10 R_{\odot}$ remain unchanged throughout their Sun-Earth propagation. The radius and magnetic field intensity of CMEs in turn were assumed to evolve from their values approximated at $\sim 10 R_{\odot}$ due to self-similar expansion \citep{subramanian2014self, vrvsnak2019heliospheric} in the course of interplanetary propagation.  In the following sections, we discuss the procedures we used to determine the model parameters. 
 
 \subsection{Estimates of the geometrical properties of FRs}
\begin{figure}[]
 \centering
 \includegraphics[width=0.5\textwidth]{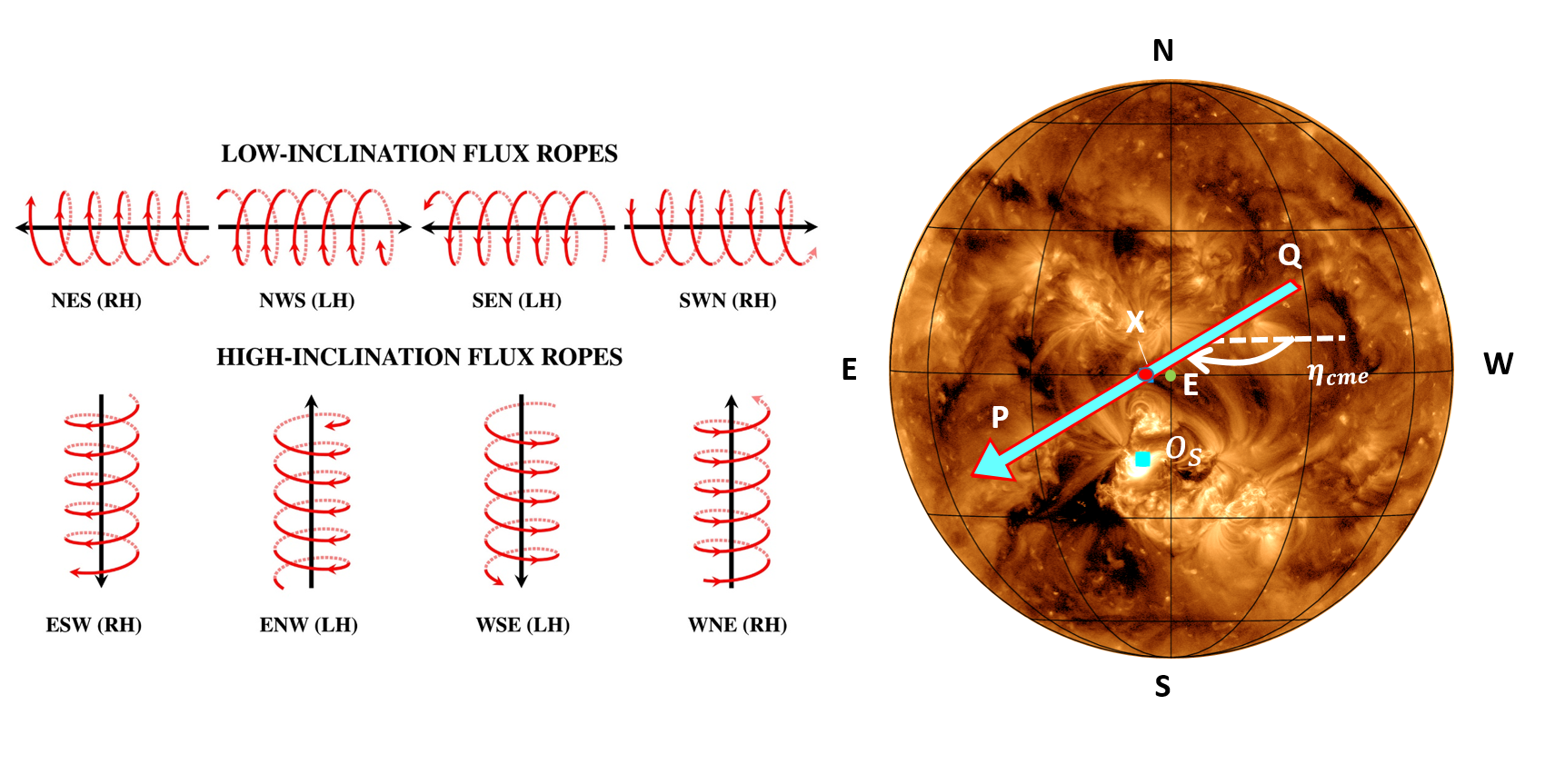}
 \caption{(a) Eight types of FR. For each type, the heilcal and axial field lines are shown in red and black, respectively. Each letter of the name of the type of FR corresponds to one of four directions, i.e. north, south, east, west, (e.g.  NES means north-east-south), and RH and LH denote right-handed and left-handed chirality, respectively. The first and last letters indicate the helical field directions, and the letter in between indicates the axial field direction of the FR. The image is adapted from \citet{palmerio2018coronal} (b) South-east directed axis of a CME (PQ) similar to the FR of event 4 shown as the cyan line projected on the solar disk. The solar source of the CME ($O_s$) is indicated by a cyan dot, and the projected location of the Earth (E) on the solar disk is denoted by a green dot. The axis has a negative value of $\eta_{cme}$ with respect to the east-west line. The $\eta_{cme}$ and $\eta_{arcade}$ associated with this FR is $-149$ and $-127$, respectively.}
\label{f5.2}
\end{figure}

 We estimated the 3D geometrical morphology of the CME and its propagation direction in the outer corona $\sim 10 --25 R_{\odot}$  by fitting the geometrical structure of CMEs using the graduated cylindrical shell (GCS) model \citep{2011ApJS..194...33T}. The CMEs are observed with the C2 and C3 coronagraphs of the Large Angle and Spectrometric Coronagraph \citep[LASCO;][]{1995SoPh..162..357B} telescope on board the Solar and Heliospheric Observatory (SOHO) and COR2 A and B of the Sun Earth Connection Coronal and Heliospheric Investigation \citep[SECCHI;][]{2008SSRv..136...67H} on board the Solar Terrestrial Relations Observatory (STEREO). By fitting the CMEs with the GCS model, we obtained the latitude ($\theta_{HG}$) and longitude ($\phi_{HG}$) of the apex of CMEs in Stonyhurst heliographic coordinates, the tilt $\eta$ ($-90 ^\circ < \eta < 90 ^\circ$) of the axis of the FR CMEs, the aspect ratio ($\kappa$) and height ($h_l$) of the CME leading edges, and the angle ($AW$) that is formed between the two legs of the CME FRs. The FR axis tilt $\eta$ is measured as counterclockwise positive from the solar west direction. The uncertainty in determining $\eta$ using the GCS is $\pm10 ^\circ$ \citep{2009SoPh..256..111T}. \citet{sarkar2020observationally} considered uncertainties of $\pm10 ^\circ$ in their $\theta_{HG}$ and $\phi_{HG}$ determinations and $\pm10\%$ when they obtained $\kappa$. Using $\theta_{HG}, \phi_{HG}$, and $\eta_{cme}$ , we formulated the FR axis and considered the Earth's location as $(\theta_{HG}, \phi_{HG}) = (0,0)$. We defined the perpendicular distance ($p$) between an MC axis and the location of a spacecraft as
 \begin{equation}
     p=\frac{|\theta_{HG}-\phi_{HG}tan (\eta)|}{\sqrt{1+tan^2 (\eta)}}.
 \end{equation}
 Using $\kappa$ , which constrains the FR expansion, we determined the initial FR radius $R_{MC}(0)$ by
 \begin{equation}
 R_{MC}(0)=\frac{height_{MC}}{1+1/\kappa},
 \end{equation}
 where $height_{MC}$ is the leading-edge height of MCs reaching Earth. Thus, it is equivalent to the Sun-Earth distance. The length ($L_{MC}$) of the FR axis at any heliocentric distance ($r$) was obtained using $AW$ by the formula $L_{MC}= (AW \times r)$, where $AW$ is in radian. In the right panel of Figure \ref{f5.2}, we show a south-east directed MC FR axis PQ projected on the solar disk with a positive $\eta_{cme}$ measured counterclockwise from the east-west direction. The Earth's location ($\theta_{HG}, \phi_{HG}) = (0,0)$) projected on the solar disk is noted by E. The CME source location ($O_s$), apex (X), and tilt ($\eta_{cme}$) are labelled in the figure.

  \subsection{Estimating the magnetic properties of the flux rope near the Sun} \label{sec5.2.3}
  The magnetic field pattern of a flux rope can be expressed in terms of an FR type \citep{angeo-16-1-1998, mulligan1998solar}. The FR type can be determined using the FR chirality: the right- or left-handed twist of the FR helical magnetic field component, the FR axis tilt, and the direction of the FR axial magnetic field. Based on these properties, FRs are classified into eight different types, including low- and high-inclination FR axes. A sketch representing the eight types of FRs is shown in the left panel of Figure \ref{f5.2}.
  
  \subsubsection{Determining the FR chirality} \label{sec5.2.3}
  To estimate the handedness of FRs, we analysed the Helioseismic Magnetic Imagers (HMI) line-of-sight (LOS) magnetograms, images obtained with the Atmospheric Imaging Assembly (AIA) on board Solar Dynamic Observatory (SDO), and H-$\alpha$ images of the solar sources of FRs. The chirality of the solar source indicates the chirality of the associated FR in the corona and farther out in interplanetary space because magnetic helicity is a conserved quantity even in magnetic reconnection \citep{2005HiA....13...85B}. The chirality is inferred by examining magnetic tongues \citep{fuentes2000counterkink, luoni2011twisted}, the dextral and sinistral nature of filament structures \citep{martin1996skew, martin2003signs}, EUV sigmoids, the skew of coronal arcades overlying the neutral lines or filament axes \citep{mcallister1995declining, martin1997predicting}, the structure of flare ribbons associated with CME FRs \citep{demoulin1996three}, and the hemispheric helicity rule \citep{angeo-16-1-1998}. Details of the chirality proxies can be found in \citet{palmerio2017determining}. \citet{palmerio2018coronal} derived the chirality of 20 ICMEs by examining the chirality proxies mentioned above and compared this with the chirality of their solar sources. The study found that 2 of the 20 ICME source regions did not follow the hemispheric helicity rule, and the in-situ flux rope chirality matched the intrinsic flux rope chirality for all 20 events. \citet{PAL2021} studied the chirality of 11 events using the chirality proxies and compared the chirality in the near-Sun CMEs and MCs at 1 AU. The study found that the solar source of 3 out of 11 events did not follow the hemispheric helicity rule. The hemispheric helicity rule may be proven as a powerful rule in the statistical sense, but the reliability of this method for individual events raises questions because only about 60-75\% of the emerging active regions follow the rule \citep{2014SSRv..186..285P}. However, it can be used to estimate the FR chirality of CMEs as a first-order approximation if CME-associated solar sources are unambiguously determined.

  \subsubsection{Determining the flux rope type of CMEs}
\begin{figure}[]
 \centering
 \includegraphics[width=0.5\textwidth]{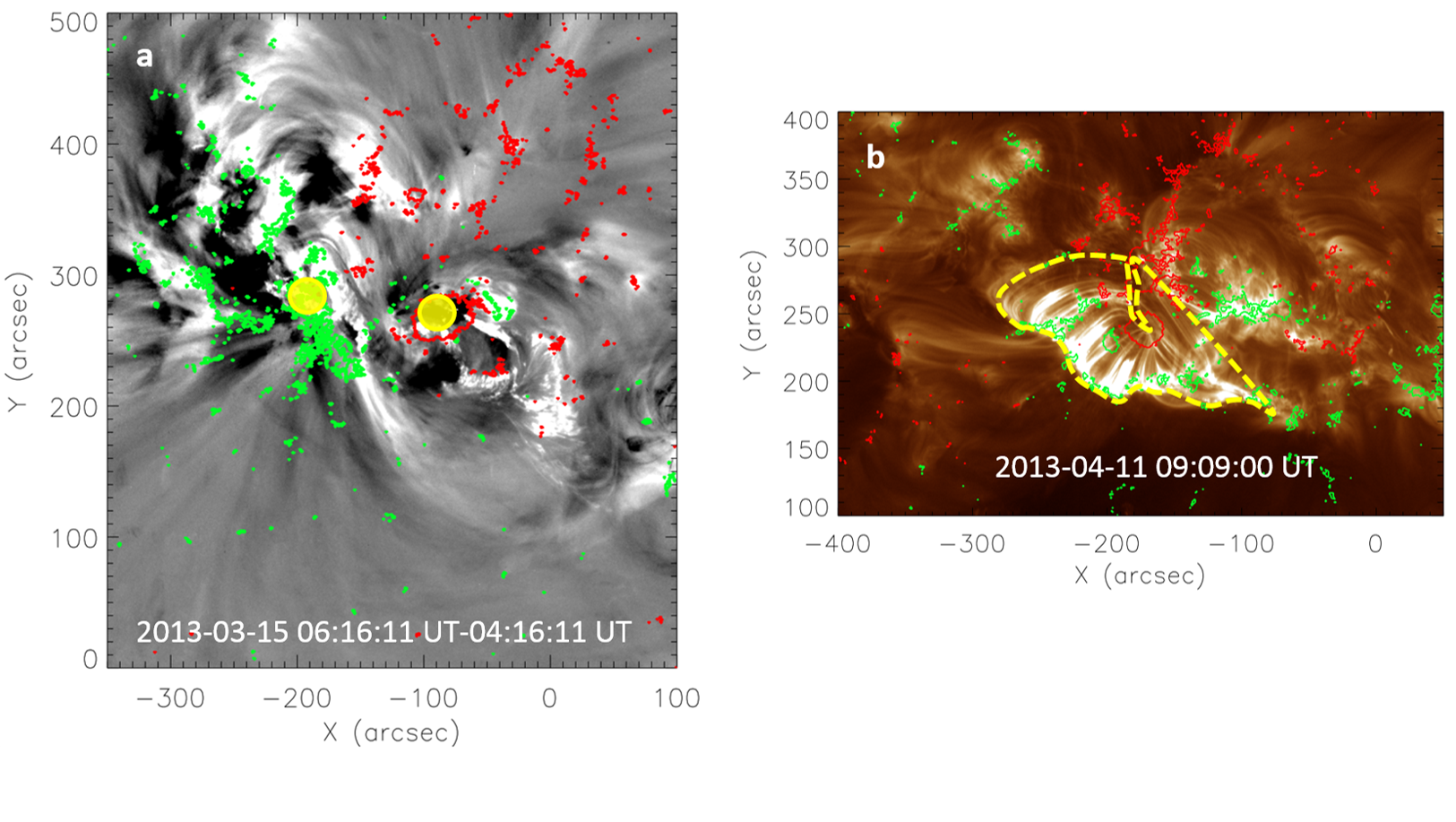}
 \caption{Determination of the magnetic FR parameters, foot points and magnetic field intensity. (a) EUV base-difference image obtained using SDO/AIA 211 $\AA$ \ observations. The regions surrounded by yellow circles denote the FR foot points. (b) observation of a PEA in SDO/AIA 193 $\AA$. The PEA foot points are indicated by dotted yellow lines. In both the images, the associated LOS magnetograms with LOS magnetic field intensity $B_{LOS} > \pm 150$ G are overplotted using green (positive magnetic field) and red (negative magnetic field) contours.}.
\label{f5.3}
\end{figure}

 The orientation of the flux-rope axis roughly follows the associated PIL \citep{2015SoPh..290..1371M} or post-eruption arcades \citep[PEAs;][]{yurchyshyn2008relationship} orientations. The flux ropes often undergo significant rotations in the lower corona during the early evolution, however, which are due to interactions with overlying skewed coronal loops \citep{lynch2009rotation}. To take the possible rotation in the corona into account, we determined the FR axis orientation ($\eta$) from the GCS at a height greater than 10 $R_\odot$ from the Sun and used this in our prediction tool. We obtained the FR foot points on the solar surface using EUV images and magnetograms. The foot points were determined by coronal dimming regions that formed during the rise period of the flux rope \citep{mandrini2005interplanetary}. We searched for EUV dimming signatures in SDO/AIA 211 $\AA$ base-difference images and overlaid an LOS magnetogram data. Thus, we obtained the magnetic polarities of FR foot points. The FR axial field is directed from positive to negative foot points. In Figure \ref{f5.3}(a) we indicate the FR foot points by yellow circles in the EUV difference image obtained using observations from SDO/AIA 211 $\AA$. The LOS magnetogram with the LOS magnetic field intensity $B_{LOS} > \pm 150$ G is overplotted on the EUV difference image using red (negative magnetic field region) and green (positive magnetic field region) contours. After obtaining chirality, the FR axis tilt, and the axial field direction, we inferred the flux rope type. Although the chirality remains the same during the FR propagation from the Sun to the Earth, the rotation of the FR axis may change the flux rope type. \citet{palmerio2018coronal} derived the FR axis tilt as the average of the orientation of PEA and PIL and found that the FR axis tilt may change ~$180 ^\circ$ from the Sun to 1 AU. However, the study obtained the near-Sun FR tilt at a lower height than we derived in our study.

  \subsubsection{Measuring the axial magnetic field intensity ($B'_{0}$) of FRs}
 To estimate the axial magnetic field strength $B_{CME}$ of CMEs near the Sun, we applied the tehcnique called flux rope from eruption data \citep[FRED]{gopalswamy2017new}, which requires the source-region reconnection flux $F_{rec}$, that is, the photospheric magnetic flux under CME associated post-eruption arcades and the length ($L_{CME}$) and radius ($R_{CME}$) of CMEs. The reconnection flux was obtained using the PEA technique discussed in \citet{2017aSoPh..292...65G} and \citet{ GOPALSWAMY2017b}. \citet{2017ApJ...851..123P} discussed the different sources of uncertainties in determining $F_{rec}$ . \citet{2021A&A...650A.176P} and \citet{PAL2021} estimated $F_{rec}$ uncertainties that emerged from errors in the PEA footpoint selection in lower corona. Because of the projection effect, $F_{rec}$ may have large uncertainties, while PEAs appear far from the solar disk centre. $F_{rec}$ is equivalent to the poloidal or azimuthal flux ($F_{pcme}$) of CME FRs \citep{2007SoPh..244...45L, 2007ApJ...659..758Q}. The poloidal flux of CMEs is conserved during their interplanetary propagation \citep{2007ApJ...659..758Q, 2014ApJ...793...53H, 2017aSoPh..292...65G,PAL2021} unless the CME flux is significantly eroded due to reconnection in the heliosphere. Thus, $B'_0$ was estimated using
  \begin{equation}
      B'_0=B_{CME}\times \frac{R_{CME}}{R_{MC}(0)}.
  \end{equation}
  
  In Figure \ref{f5.3}(b), a PEA region is indicated by the dashed yellow line in the SDO/AIA 193 $\AA$ image. The positive and negative magnetic field regions are shown by green and red contours, respectively.

 \subsection{Estimating the arrival time and transit speed of CMEs }
 Along with the magnetic profile, we estimated the CME arrival time and transit speed at 1 AU using the web tool of the probabilistic drag-based ensemble model \citep[DBEMv3;][]{vcalogovic2021probabilistic}. This is an upgraded version of the 2D flattening cone drag-based model \citep[DBM;][]{vrvsnak2010role,2013SoPh..285..295V,vzic2015heliospheric}, which combines the cone geometry describing the propagation of CME leading edge and the concept of aerodynamic drag on the interplanetary propagation of CMEs. The drag-based ensemble model \citep[DBEM;][]{2018ApJ...854..180D} produces possible distributions of CME arrival information by using an ensemble of $n$ measurements of the same CME, where the CME ensemble is produced by the observer. The model assumes the CME to be a cone structure with a semi-circle leading-edge spanning over its angular width where the structure flattens with the CME's interplanetary evolution \citep{vzic2015heliospheric}. It considers the solar wind speed ($V_{sw}$) and drag parameter ($\gamma$) to be constant beyond the distance of 15 $R_\odot$. This is because beyond 15 $R_\odot$ CMEs propagate through an isotropic solar wind that has a constant velocity. Moreover, the rate of the fall-off of solar wind density is similar to the rate of the self-similar expansion of CMEs \citep{vrvsnak2013propagation, vzic2015heliospheric}. The DBEMv3 is an advanced form of the DBEM in which the model generates ensemble members based on the observational inputs and their uncertainties that are provided to the model. The inputs to the model are the initial CME speed $V_{CME}$ with the uncertainty $\pm\Delta V_{CME}$, the half angular width $\lambda$ projected on the plane-of-sky with the uncertainty $\pm\Delta \lambda$, the propagation longitude $\phi_{HG}$ with the uncertainty $\pm\Delta \phi_{HG}$ at a specific radial distance $R_0$, the CME arrival time $t_{launch}$ at $R_0$ with the uncertainty $\pm\Delta t_{launch}$ along with the radial speed of the solar wind $V_{sw}\pm\Delta V_{sw}$ and the drag parameter $\gamma\pm\Delta \gamma$. DBEMv3 is available on the Hvar Observatory website as an online tool \footnote{\url{http://phyk039240.uni-graz.at:8080/DBEMv3/dbem.php}} . It is a product of the European Space Agency (ESA) space situational awareness (SSA).\par
 
 We assumed that at 1 AU, the plasma propagation speed ($V_{pro}$) within CMEs is almost equal to their average Sun-Earth transition speed $V_{tr}$ \citep{lepping2008estimates}. To prepare the inputs to DBEMv3, we used the CME parameters obtained from the GCS fitting results, where the GCS model was fitted to CMEs at a height more than 10 $R_{\odot}$. We derived the initial CME speed $V_{CME}$ and its arrival time at $R_0 = 21.5 R_\odot$ by least-square fitting its height-time profile. Following \citet{vcalogovic2021probabilistic}, the uncertainties $\Delta t_{launch}$, $\Delta\lambda$, and $\Delta \phi_{HG}$ were set to $\pm 30$ min, $\pm 10\%,$ and $\pm5^\circ$, respectively. For each CME, the drag parameter $\gamma$ was selected based on their speed. The values of $\gamma$ are based on empirical data  \citep{vrvsnak2013propagation, vrvsnak2014heliospheric, vzic2015heliospheric}. For CMEs with $V_{CME}< 600$ km/s, $\gamma$ was set to $0.5 \times 10^{-7}\pm0.1$ km$^{-1}$, for 600 km/s $< V_{CME}<1000$ km/s, $\gamma$ was set to $0.2\times 10^{-7}\pm0.075$ km$^{-1}$ , and for $ V_{CME}>1000$ km/s, $\gamma$ was set to $0.1\times 10^{-7}\pm0.05$ km$^{-1}$ \citep{vcalogovic2021probabilistic}. \par 
 To estimate the ambient solar wind speed ($V_{sw}$), we followed the  empirical solar wind forecast \citep[ESWF;][]{vrvsnak2007transit, vrvsnak2007coronal,rotter2012relation,rotter2015real,reiss2016verification} processes that monitor fractional areas that are covered by coronal holes close to the central meridian region. We followed an algorithm based on an empirical relation that links the area of coronal holes that appeared close to the central meridian ($\pm 10 ^{\circ}$) and solar wind speed. The empirical relation follows the equation $V_{sw}(t)=c_0+c_1A(t-\delta t)$, where $A$ is the fractional coronal hole area. We shifted coronal hole area time series with a time lag $\delta t$ to determine the $V_{sw}$ at time $t$. \citet{vrvsnak2007coronal} studied this empirical relation during the period DOY 25 – 125 in 2005 and found $\delta t= 4$ days, $c_0 = 350$ km/s, and $c_1= 900$. They found that the average relative difference between the predicted and observed peak solar wind speed values is $\pm10 \%$. Therefore, we considered $\Delta V_{sw} = \pm10\%$

\subsection{Coordinate conversion of the magnetic field vectors}
At 1 AU, the inclination angle ($\theta_{MC}$) of MCs is considered to be equivalent to the $\eta$ of the associated CMEs, and the azimuthal angle ($\phi_{MC}$) of MCs is determined using CME propagation longitudes ($\phi_{HG}$) obtained at 10 $R_\odot$. In order to express $B_{az}$, $B_{ax}$ , and $B_{\rho}$ in the Geocentric Solar Ecliptic (GSE) coordinate system (a Cartesian coordinate system where $\hat{z}$ is perpendicular to the Sun-Earth plane, and $\hat{x}$ is parallel to the Sun-Earth line and positive toward the Sun) that is commonly used to represent the magnetic field vectors of ICMEs at 1 AU, we transformed the field vectors from the local cylindrical to the Cartesian coordinate system. First, $B_{az}$, $B_{ax}$, and $B_{\rho}$ were converted into $B_{x,cl}, B_{y,cl}$ , and $B_{z,cl}$ , which are in the local Cartesian coordinate ($\hat{x}_{cl}, \hat{y}_{cl}$, $\hat{z}_{cl}$) system originating at the MC axis. Finally, using $\theta_{MC}$ and $\phi_{MC}$, the magnetic field vectors $B_{x,cl}$, $B_{y,cl}$ , and $B_{z,cl}$ were transformed into $B_{x}$, $B_{y}$ , and $B_{z}$.  \par
\begin{figure}[]
 \centering
 \includegraphics[width=0.5\textwidth]{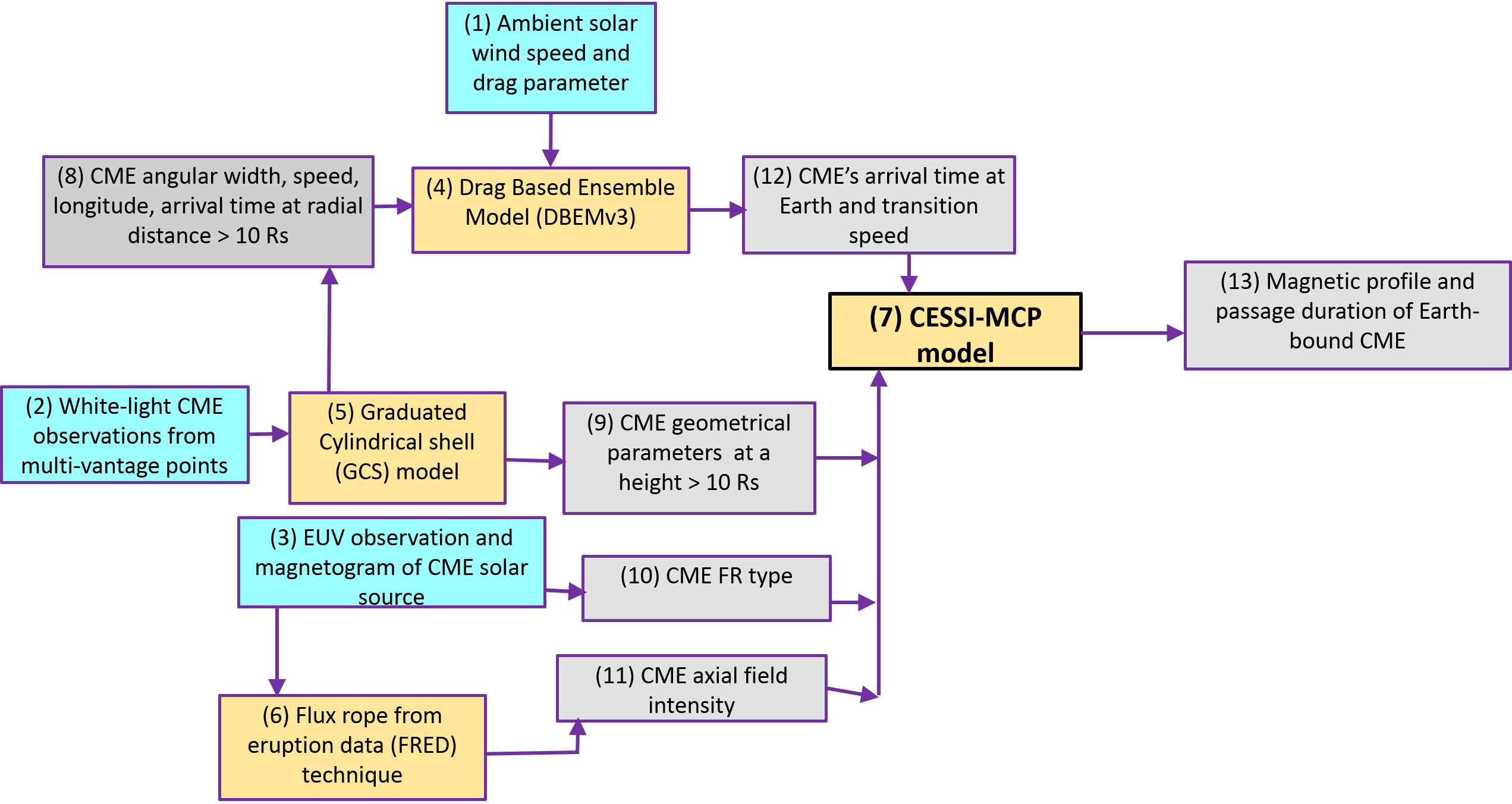}
 \caption{Block diagram representing the steps involved in our analytical approach to predict Earth-bound CME magnetic vectors and passage duration. The cyan blocks 1--3 contain near-Sun remote observations that were used as inputs. The yellow blocks 4--7 indicate the models and techniques involved in our approach. The grey blocks 8--13 indicate the outputs.}
\label{f5.4}
\end{figure}

 
In Figure \ref{f5.4} we present our MC prediction approach using a block diagram. The yellow blocks represent the models and techniques used in our approach, cyan blocks indicate the model input parameters, and the grey blocks indicate the outputs derived from the models and techniques used here. In Table\ref{summary} we provide the models and techniques, instruments, model inputs, and outputs used in this study.

\begin {table*}[!h]

\vspace{-4mm}
\centering
\begin{threeparttable}
\caption {Summary of the required models and techniques, satellite instruments, and inputs.   }

\begin{tabular}{ p{4cm} p{4cm} p{2cm} p{2cm} }
 \hline

 Models/Techniques & Instruments & Inputs& Outputs\\
 
  (1)&(2)&(3)&(4)\\
  \hline
  GCS \tnote{a} &SOHO/LASCO C2, C3, STEREO/SECCHI/COR2 A\&B& Coronagraphs&$\theta_{HG}$, $\phi_{HG}$,  $\eta$, $\kappa$, $h_l$, $AW$, $V_{CME}$\\
  
  DBEMv3 \tnote{b}&& $V_{sw}$, $\gamma$, $\phi_{HG}$,  $V_{CME}$, $\lambda$, $t_{launch}$&$V_{tr}$, $t_{ar}$\\
  
  FRED\tnote{c}  & SDO/HMI, SDO/AIA $193\AA$& EUV images and magnetograms &$B_{CME}$\\
  MCP &SDO/AIA 131$\AA$, 171$\AA$, 1600$\AA$, 304$\AA$, 211$\AA$, H-$\alpha$ imagery, SDO/HMI&$\theta_{HG}$, $\phi_{HG}$,  $\eta$, $\kappa$, $h_l$, $AW$, $B_{CME}$, $V_{tr}$, $t_{ar}$ , flux rope type & $B_x,B_y,B_z$, MC passage duration\\

  \hline
  \\

\end{tabular}
\label{summary}
\vspace{-4mm}
\begin{tablenotes}
\item[a] \citet{2011ApJS..194...33T}
\item[b] \citet{vcalogovic2021probabilistic, 2018ApJ...854..180D}
\item[c]\citet{gopalswamy2017new}
\end{tablenotes}
\end{threeparttable}
\end{table*} 

\section{Results: Model validation using observed MC events} \label{sec5.3}
As a proof of concept, we validated our model by investigating ten  Earth-directed MCs appearing as FRs in near-Sun and in-situ regions. The MCs had clearly identified solar sources. In the near-Earth region (L1 Lagrangian point), MCs are observed using the Magnetic Field Experiment (MAG) instrument of the Advanced Composition Explorer (ACE) spacecraft. The events were selected from the Richardson \& Cane ICME catalogue \footnote{\url{http://www.srl.caltech.edu/ACE/ASC/DATA/level3/icmetable2.html}} \citep{Richardson2010}. The front and rear boundaries of the MCs were verified manually such that at 1 AU, they maintained the MC properties suggested by \citet{1981JGR....86.6673B} throughout their interval, and their associated CMEs appear as isolated magnetic structures in near-Sun observations. \par
\subsection{Preparation of model inputs}
We manually identified each of the MC associated CMEs following \citet{2007JGRA..11212103Z} and \citet{2017ApJ...851..123P} and located their solar sources using their coronal signatures observed in SDO/AIA. We obtained the geometrical parameters  of the CMEs, $\theta_{HG}$,  $\phi_{HG}$, $AW$, $\eta$, and $\kappa$ at a height $h_l > 10 R_{\odot}$ and tabulate them in Columns 5-10 of Table \ref{t5.1}, respectively. The GCS fitting to FRs associated with CMEs 4 and 7 can be found in Figure 5 and 1 of \citet{2017ApJ...851..123P} and \citet{pal2018dependence}, respectively. We defined the CME initiation time ($CME_{start}$) as the moment at which the CMEs were first identified in the SOHO/LASCO C2 field of view. In Columns 1 and 2, the event numbers (Ev no.) and $CME_{start}$  are listed, respectively. Columns 3 and 4 contain the start ($MC_{start}$) and end time ($MC_{end}$) of MCs adapted from the Richardson \& Cane ICME catalogue.  \par

\begin {table*}[!h]

\vspace{-4mm}
\begin{center}
\caption {Near-Sun observations of $\theta_{HG}$, $\phi_{HG}$, $\eta$, $\kappa$, $h_l$, $AW$, $CME_{start}$, and associated $MC_{start}$ and $MC_{end}$. }
\centering

\begin{tabular}{ p{0.5cm} p{1.5cm} p{1.5cm} p{1.5cm} p{1cm} p{0.90cm} p{1.0cm} p{.85cm} p{.9cm} p{0.8cm}p{0.8cm}}
 \hline
\centering
 Ev no.&$CME_{start}$ (UT) &$MC_{start}$ (UT)& $MC_{end}$ (UT) &$Dur_{obs}$ (hr)&$\theta_{HG}$  $(^{\circ})$ &$\phi_{HG}$ $(^{\circ})$&$\eta$ $ (^{\circ})$&$AW $ $(^{\circ})$&$\kappa$&$h_l$ $(R_{\odot}$) \\
  (1)&(2)&(3)&(4)&(5)&(6)&(7)&(8)&(9)&(10)&(11)\\
  \hline
1&2010/05/24 14:06:00&2010/05/28 20:46:00&2010/05/29 16:27:00&20&0&     5.3&$-53.3$&    36&0.22&11.4\\
2&2011/06/02 08:12:00&2011/06/05 01:50:00& 2011/06/05 19:00:00&17& $-7.8$&      $-11.8$&        55.3&   34&     0.15&14.7\\
3& 2012/02/10 20:00:00& 2012/02/14 20:24:00& 2012/02/16 05:34:00&33& 28&$       -23$&$  -72$&   50&     0.23&13.5\\
4& 2012/06/14 14:12:00 &2012/06/16 22:00:00&    2012/06/17 14:00:00& 16&0&      $-5$&   30.7&   76&     0.30&15\\
5 &2012/07/12 16:48:00& 2012/07/15 06:00:00&    2012/07/17 05:00:00&47&$-8$&    14      &53.1&  60      &0.66&14.1      \\
6 &2012/11/09 15:12:00 &2012/11/13 09:44:00&    2012/11/14 02:49:00& 24&$2.8$&  $-4$&   $-2$&   $36$&   0.20&12.4       \\
7 &2013/03/15 07:12:00 &2013/03/17 14:00:00&    2013/03/18 00:45:00& 11&$-6.5$& $-10$   &$-74.4$        &51&    0.27&18\\
8 &2013/04/11 07:24:00 &2013/04/14 16:41:00&    2013/04/15 20:49:00&28&$ -5.5$&  $-15$&  68.2&   74&     0.24&20.5\\
9 &2013/06/02 20:00:00 &2013/06/06 14:23:00&    2013/06/08 00:00:00& 33&$-1.7$& 7&      75.5&   36&     0.21    &12.14\\ 
10 &2013/07/09 15:12:00&2013/07/13 04:39:00&    2013/07/15 00:00:00& 43&2       &3&     $-37.5$&        36&     0.36&13.4\\
\hline
\end{tabular}
\label{t5.1}
\end{center}
\vspace{-4mm}
\end{table*}
Using $V_{sw}$ (derived using the empirical relation between the coronal hole area and the solar wind speed), the drag parameter, the deprojected CME velocity, the longitude and projected angular width and their uncertainties as input, DBEMv3 estimates the probability ($p_{tar}$) of the CME arrival at Earth, the arrival time, and the arrival speed distributions. In Columns 2, 3, and 4 of Table \ref{t5.2}, we provide the extrapolated deprojected speed of CMEs at 21.5 $R_\odot$, $V_{sw}$ during CME propagation, and $p_{tar}$, respectively. The median $t_{ar}$ of arrival time distribution with 95\% confidence intervals ($t_{ar,LCI}  < t_{ar} < t_{ar,HCI}$) is listed in Column 5 of Table \ref{t5.2}. We computed the CME transit speed $V_{tr}$ using the Sun-Earth distance and $t_{ar}$. The error in the arrival time prediction was obtained from $t_{err} = t_{ar} - MC_{start}$. We present $V_{tr}$ and $t_{err}$ in Columns 6 and 7. For the ten events considered here, we find a mean error ($ME$) and mean absolute error ($MAE$) in the prediction of the CME arrival time as $ME\sim-4.75$ hours and $\sim$6.3 hours, respectively. A negative $ME$ value indicates that CMEs are predicted to arrive earlier at L1 than they are observed in most cases. This may be caused by an overestimation of $t_{launch}$ or by the physical limitations in the DBM model. \citet{vcalogovic2021probabilistic} applied DBEMv3 on 146 CME-ICME pairs to evaluate the performance of DBEMv3 and obtained an $ME$ of $-11.3$ hours and an $MAE$ of 17.3 hours. They discussed the requirement of fine-tuning the DBEM input parameters for extreme CMEs due to their specific CME properties and the complex heliospheric conditions through which they propagate. The study noted that although a higher value of $\gamma$ may improve the travel time predictions for fast CMEs, it increases the prediction errors of the CME arrival speed. \citet{dumbovic2018drag} evaluated DBEM on 35 CME-ICME pairs that were compiled and analysed by \citet{mays2015ensemble} and found that the model errors were comparable to those of the ensemble WSA-ENLIL+Cone model \citep{odstrcil2004numerical}. However, DBEM does not perform well during solar maximum \citep{vrvsnak2013propagation}; when the heliospheric environment becomes complex, CME-CME interaction becomes inevitable \citep{2020ApJ...899...47R}, and CMEs frequently propagate through high-speed streams \citep{2010A&A...512A..43V}. These conditions significantly influence the two important DBEM input parameters, the drag parameter and the background solar wind speed.   
\par

\begin {table*}[!h]

\begin{center}
\caption { $V_{CME}$, $V_{SW}$, $p_{tar}$, predicted arrival time range ($t_{ar,LCI} <t_{ar}<t_{ar,HCI}$), $V_{tr}$ and the difference $t_{err}$ between the observed and predicted median arrival times.  }
\centering

\begin{tabular}{ p{0.5cm} p{1.2cm} p{1cm} p{1cm}p{8.5cm} p{1cm} p{1cm} }
 \hline

 Ev no.&$V_{CME}$&$V_{SW}$&$p_{tar}$ &$t_{ar,LCI} <t_{ar}<t_{ar,HCI}$&$V_{tr}$ &$t_{err}$ \\
 &km/s&km/s&\%&UT&km/s&Hr\\
  (1)&(2)&(3)&(4)&(5)&(6)&(7)\\
  \hline
1& 562$\pm$60&350$\pm$35 &100&228-05-2010 05:57 < 28-05-2010 10:05 < 28-05-2010 14:42&450&$-10.7$\\
2&937$\pm$20&356$\pm$36&100&04-06-2011 20:09 < 05-06-2011 00:11 < 05-06-2011 04:17&646&$-1.7$\\
3&658$\pm$13&352$\pm$35&51&14-02-2012 05:05 < 14-02-2012 09:41 < 14-02-2012 13:35&482&$-10.7$\\
4&1020$\pm$60&350$\pm$35&100&16-06-2012 13:57 < 16-06-2012 17:31 < 16-06-2012 21:00&806&$-4.5$\\
5&1000$\pm$200&363$\pm$36&100&15-07-2012 00:20 < 15-07-2012 05:42 < 15-07-2012 11:15&679&$-0.3$\\
6&611$\pm$24&350$\pm$35&100&12-11-2012 18:22 < 12-11-2012 21:22 < 13-11-2012 00:38&529&$-11$\\
7&1160$\pm$28&355$\pm$35&100&17-03-2013 02:47 < 17-03-2013 06:20 < 17-03-2013 09:59&877&$-7.7$\\
8&700$\pm$24&371$\pm$37&99.8&14-04-2013 07:55 < 14-04-2013 11:53 < 14-04-2013 16:40&599&$-4.8$\\
9&500$\pm$70&350$\pm$35&100&06-06-2013 16:42 < 06-06-2013 22:17 < 07-06-2013 04:47&383&7.9\\
10&610$\pm$30&350$\pm$35&100&12-07-2013 21:28 < 13-07-2013 00:37 < 13-07-2013 04:04&508&$-4$\\
\hline

\end{tabular}

\label{t5.2}
\end{center}
\end{table*}

To determine FR types, we obtained their chirality, axis orientations, and axial magnetic field directions using remote observations as described before. The multi-wavelength proxies mentioned in Section \ref{sec5.2.3} were examined for all MCs to infer their chirality. We converted PEA tilt and FR axis tilt $\eta$ into the orientation angles $\eta_{arcade}$ and $\eta_{cme}$ respectively, which lie within the range [$-180 ^\circ$, 180$^\circ$]. The angles were measured from the solar west direction, counterclockwise for positive and clockwise for negative values. They were derived by taking the axial field direction into account, which was estimated using coronal dimming information. \citet{yurchyshyn2008relationship} studied the relation between PEA angles and CME directions of 25 FR events and found that for majority of events, the difference between the angles remains smaller than 45$^\circ$. In Table \ref{chirality} we summarise the near-Sun FR magnetic properties of ten events. Column 1 shows the event numbers, Column 2 shows the flux-rope chirality, where $+1$ stands for right-handedness and $-1$ represents left-handedness. In Columns 3, 4, and 5 we present $\eta_{arcade}$, $\eta_{cme}$ and their difference $\eta_{diff}$, where a positive value of $\eta_{diff}$ represents the rotation of the CME axis counterclockwise with respect to the PEA tilt. 
  
  \begin {table*}[!h]

\vspace{-4mm}
\begin{center}
\caption {CME near-Sun magnetic properties, PEA tilt ($\eta_{arcade}$), CME axis orientation ($\eta_{cme}$), their differences ($\eta_{diff}$), near-Sun FR type (type$_{ns}$), and axial magnetic field intensity (F$_{pcme}$).   }
\centering

\begin{tabular}{ p{1cm} p{1.2cm} p{1.2cm} p{1.2cm} p{1.2cm} p{1.2cm} p{1.5cm} }
 \hline

 Ev no.&Chirality&$\eta_{arcade}$&$\eta_{cme}$&$\eta_{diff}$&type$_{ns}$&F$_{pcme}$  \\
 &&$(^{\circ})$&$(^{\circ})$&$(^{\circ})$&&($10^{21}$ Mx)\\
 
  (1)&(2)&(3)&(4)&(5)&(6)&(7)\\
  \hline
  
  1&LH&$-19.3$&$-53.3$&$-34$&WSE&2.15\\
  2&RH&46.2&55.3&9&WNE&1.81\\
  3&RH&$110.2$&72&$-38.2$&ESW&2\\
  4&RH&$-127.4$&$-149.3$&$-22$&NES&8.45\\
  5&RH&$-151$&$-127$&24&ESW&14.10\\
  6&RH&$-172$&178&$-6$&NES&2.47\\
  7&RH&$123$&105.6&$-17.4$&ESW&4.10\\
  8&LH&$-61.5$&$-111.8$&$-50.3$&ENW&3.72\\
  9&LH&$137.7$&75.5&$-62.2$&WSE&1.75\\
  10&LH&$-39.2$&$-37.5$&1.7&NWS&3.50\\
  \hline

\end{tabular}
\label{chirality}
\end{center}
\vspace{-4mm}
\end{table*} 
   
   As $\eta_{cme}$ is measured at a coronal height ($\geq$ 10 R$_\odot$ ) greater than that where the $\eta_{arcade}$ is measured, we used $\eta_{cme}$ as the final value of the FR axial orientation. Combining $\eta_{cme}$, chirality, and FR axis direction, we estimated the type of CME FRs, $type_{ns}$ , and list it in Column 6 of Table \ref{chirality}. Finally, the axial magnetic field intensities of the associated CMEs were derived using $F_{pcme}$ and the FR geometrical parameters. Column 7 of Table \ref{chirality} shows $F_{pcme}$ of near-Sun FRs.
\par

\subsection{Model outputs} 

Using the near-Sun CME observations as input to the constant-$\alpha$ force-free cylindrical FR model that expands self-similarly in radial directions, we estimated the magnetic field vectors of the associated MCs intersecting the spacecraft at 1 AU. To incorporate the ambiguities involved in measurements of propagation direction, inclination, and size of the CMEs, we used the uncertainty range of these parameters as input to our model. We prepared ten different random input sets of each MC where the input parameter values are within $\pm 10^\circ$ of measured propagation direction and $\eta_{cme}$, and $\pm 10\%$ of estimated $\kappa$ value. The magnetic field vectors were derived using each of the input sets. Thus, we obtained ten different magnetic profiles for every event and measured the root-mean-square (rms) differences between the observed and predicted magnetic vectors. The normalised rms difference ($\Delta_{rms}$) was calculated using the ratio of $\delta B$ and $B^o_{max}$, where $B^o_{max}$ is the maximum observed magnetic field intensity, and $\delta B$ is defined by
\begin{equation}
    \delta B=\sqrt{\frac{\sum_{i}(\mathbf{B^o(t_i)}-\mathbf{B^p(t_i)})^2}{N}}.
\end{equation}

\begin {table*}[!h]
\begin{center}
\caption {Latitude and longitude of MC axes derived from observed and predicted magnetic field vectors, predicted $Y^m_0$, $\Delta_{rms}$, type$_{ne}$,  $C_{or}$ , and minimum and maximum values of the predicted duration ranges ($Dur_{pred,min}-Dur_{pred,max}$). }
\centering

\begin{tabular}{ p{1cm} p{1.2cm} p{1cm} p{1.2cm} p{1cm} p{1cm} p{1.2cm} p{1.2cm}p{0.8cm} p{3.5cm}}
 \hline

 Ev no.&$\theta^{mva}_{MC} $ &$\phi^{mva}_{MC} (^{\circ})$&$\theta^m_{MC} (^{\circ})$ &$\phi^m_{MC} (^{\circ})$& $Y^m_0$ &$\Delta^m_{rms}$ &$type_{ne}$ &$C_{or}$&$Dur_{pred,min}$-$Dur_{pred,max}$\\
 &$(^{\circ})$&$(^{\circ})$&$(^{\circ})$&$(^{\circ})$&&&&&(Hr)\\
  (1)&(2)&(3)&(4)&(5)&(6)&(7)&(9)&(10)&(11)\\
  \hline
1 &$-69$&244.5& $-$60& 273&0.5& 0.07 &WSE (LH)&$y$&27 - 41\\
2&45.8&193&59&      281&    $-$0.62&     0.22&WNE (RH)&$y$&14 - 20\\
3&$-10.5$&271&-64&  307&    -0.2&  0.1&ESW (RH)&$y$&33 - 43\\

4&$-7.5$&102.4&$-30.6$& 84.1&$-0.15$&     0.08&NES (RH)&$y$&19 - 22\\
5&$-76.3$&183.9&$-62.7$&      151&     0.73&    0.07 &ESW (RH)&$y$&38 - 50\\

6&8&83.4&3.3&      84.7&$   -0.1$&    0.07&NES (RH)&$y$&16 - 33\\

7&$-15.9$&16&$-72$&      297.5&     0.72&      0.2&SWN (RH)&$n$&20 - 25\\

8&59.8&337
&$74.7$ &      304&    $-0.54$& 0.1&ENW (LH)&$y$&10 - 24\\

9&$-81.7$&193.2&$-70.9$&      97.24&     0.28&        0.06&WSE (LH)&$y$&28 - 57\\

10&$-9$&284&$-36$&     263.7&    $-0.53$&    0.05&NWS (LH)&$y$&35 - 57\\
\hline
\end{tabular}
\label{t5.4}
\end{center}
\vspace{-4mm}
\end{table*}

\begin{figure}[]
  \centering
 
                \includegraphics[width=0.5\textwidth]{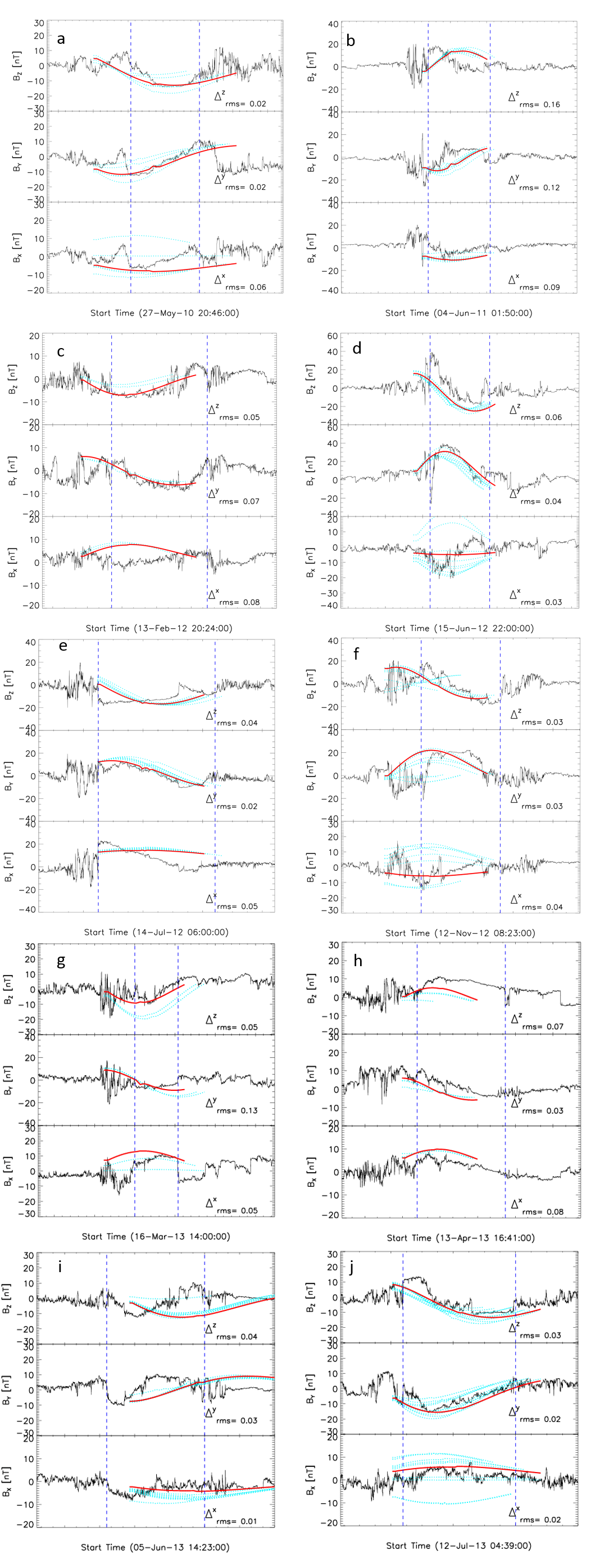}
                \caption{Magnetic vectors (in black) of ten MCs as observed by the ACE spacecraft. The red curves represent the predicted magnetic vectors that match the observed magnetic vectors best. The cyan dotted curves show the uncertainty in predictions. The vertical dashed blue lines denote the start and end time of MCs. The rms differences $\Delta^x_{rms}, \Delta^y_{rms}$, and $\Delta^z_{rms}$ between observed and predicted magnetic vectors $B_x$, $B_y$ , and $B_z$ are labelled in the plot.} 
 \label{f5.5}
\end{figure}

Here $\mathbf{B^o(t_i)}$ and $\mathbf{B^p(t_i)}$ are the observed and predicted magnetic field vectors, respectively, and $i= 1, 2, 3...N,$ with N being the total number of data points in the predicted magnetic vectors. The observed magnetic field vector was binned with a bin size=($\frac{MC_{end}-MC_{start}}{N}$). We obtained the MC axis orientations ($\theta^m_{MC}$, $\phi^m_{MC}$) and impact parameters corresponding to the predicted MC magnetic profiles with a minimum value of $\Delta_{rms}$. The $\Delta_{rms}$ was estimated for $B_x$, $B_y$ , and $B_z$ separately and is represented by $\Delta^x_{rms}, \Delta^y_{rms}$ , and $\Delta^z_{rms}$, respectively. In Figure \ref{f5.5} we display the predicted magnetic vectors obtained from the model along with the in-situ data measured at L1 by the ACE/MAG instrument for ten MCs. The observed solar wind magnetic field vectors are shown in black, and the red curves overplotted on them during the MC intervals (indicated by dashed vertical blue lines) represent the predicted fields with a minimum value of $\Delta_{rms}$. The uncertainties in the predictions resulting from errors in the input estimates are shown using dotted cyan curves. The latitude, longitude, and normalised impact parameter $Y^m_0= \frac{p}{R_{MC}(0)}$ corresponding to the minimum value of $\Delta_{rms}$, that is, $\Delta^m_{rms}$ of individual cases are denoted by $\theta^m_{MC}$,  $\phi^m_{MC}$ , and $Y^m_0$, respectively. We applied a minimum variance analysis \citep{1967JGR....72..171S} to the in-situ measurements and estimated the orientation (latitude $\theta^{mva}_{MC}$, and longitude $\phi^{mva}_{MC}$) of MCs at 1 AU, in order to compare the predicted and observed orientations. In Columns 2 and 3 of Table \ref{t5.4}, we present $\theta^{mva}_{MC}$ and $\phi^{mva}_{MC}$ values. We list the values of $\theta^m_{MC}$,  $\phi^m_{MC}$, $Y^m_0$, and $\Delta^m_{rms}$ of ten MCs in Columns 4, 5, 6, and 7 of Table \ref{t5.4}, respectively. The magnetic type of the MCs ($type_{ne}$) as observed by ACE is noted in Column 9. To compare the magnetic field orientation in predicted and observed FRs at 1 AU, we used the parameter $C_{or}$ in Column 8. Here $y$ and $n$ indicate a match and mismatch in field line orientation of near-Sun and near-Earth FRs, respectively. The event numbers (Ev no.) are listed in Column 1. Using $R_{MC}$ and the plasma propagation speed inside FRs, we estimate a range of predicted duration values ($Dur_{pred}$) for each FR at 1 AU. We list the minimum ($Dur_{pred,min}$) and maximum ($Dur_{pred,max}$) values of $Dur_{pred}$ range in Column 11 of Table \ref{t5.4}. The observed MC duration $Dur_{obs}$ values are listed in Column 5 of Table \ref{t5.1}. When the minimum and maximum values of $Dur_{pred}$ and $Dur_{obs}$ are compared, the $Dur_{obs}$ of events 1, 4, and 7 does not fall in the $Dur_{pred}$ range and is shorter than $Dur_{pred,min}$ by 8, 3, and 10 hours, respectively. We note that the overestimation in $Dur_{pred}$ mostly results from  $t_{err}$ and errors in FR radius estimates.

\section{Discussion} \label{sec5.4}

The modelling framework presented here allows prior estimates of the magnetic field profile of MCs at 1 AU, their arrival time, average speed while crossing the Earth, and duration of passage. It thus provides comprehensive intelligence about impending space weather events.
 
The approach constrains CME FRs using remote solar observations and takes the radial expansion of MCs into account. It assumes MCs to expand self-similarly during their Sun-Earth propagation. The geometric and kinematic parameters of MCs are constrained using the GCS fitting to the white-light coronagraph images of associated CMEs at a height greater than 10 $R_\odot$, while their magnetic parameters are constrained using remote observations of their solar sources. When the near-Sun CME parametrisation is performed, our analytical model takes only a few seconds to predict the profiles and estimates the approximate duration of Earth-directed CMEs. The near-real time data ($\sim 6$ hour before the current time) from the LASCO and SECCHI coronagraphs are available in SOHO \footnote{\url{https://sohoftp.nascom.nasa.gov/qkl/lasco/quicklook/level_05}} and STEREO Science Center \footnote{\url{https://stereo-ssc.nascom.nasa.gov/data/beacon/}}, respectively. Furthermore, the near-real time observations ($ \text{about one}$ hour before the current time) of the solar atmosphere and the photospheric magnetic field from AIA and HMI on board the SDO spacecraft are available in Atmospheric Imaging Assembly \footnote{\url{https://sdowww.lmsal.com/suntoday_v2/}} and Joint Science Operations Center (JSOC) \footnote{\url{https://jsoc.stanford.edu/data/hmi/fits/}}, respectively. Using these resources, the model is able to predict the properties of CMEs reaching the Earth within 6 hours of their initiation from the Sun. Typically, CMEs may take 15 hours to several days to reach Earth after leaving the Sun. However, the near-real time observations do not provide science-quality data. To acquire the desired lead time in forecasting the CME geo-effectiveness, beacon data can be used. To predict the arrival of FRs at Earth, the drag-based ensemble model is applied. \par

We applied the CESSI-MCP modelling framework to predict the magnetic profile of ten Earth-directed CMEs with clear in-situ flux rope signatures at 1 AU. They evolved as isolated magnetic structures from the Sun to the Earth, had precisely identifiable solar origins located near the centre of the solar disk (within 40$^\circ$ from the central meridian) and had available remote observations of solar sources. The flux rope of ICMEs (or, in other words, the MCs) may be embedded in extended ICME intervals, which may result in an ambiguous FR identification. \citet{2013AnGeo..31.1251K} discussed the probable reasons for significant differences between the boundaries of MC and ICMEs. The CME-CME interaction may disturb the ambient condition during the CME propagation and distort the MC boundaries. Moreover, if CMEs originate from active regions with a complex magnetic topology, the ICMEs may exhibit distorted magnetic and plasma structures at their front and rear. Interactions between MCs and IMFs may sometimes result in the accumulation of magnetic field lines at the front and/or rear of FRs and cause deflection and/or rotation in CMEs. In these cases, our modelling framework does not perform well because its input parameters are constrained only by near-Sun observations, and it assumes that flux ropes are non-interacting, force-free cylindrical structures undergoing self-similar expansion.        \par

\begin{figure}[]
 \centering
 \includegraphics[width=0.5\textwidth]{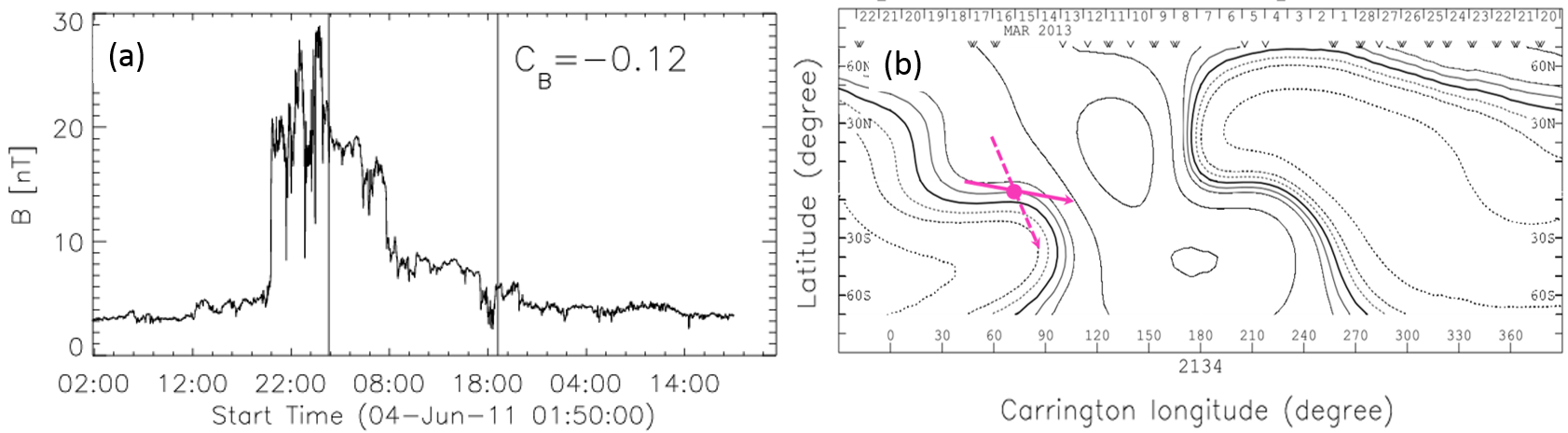}
 \caption{(a) Asymmetry in in-situ magnetic field intensity associated with event 2 MC. The FR asymmetry parameter $C_B$ is lablled in the figure. The vertical lines represent the start and end times of the MC. (b) coronal map during CR 2134 collected from the Wilcox solar observatory. The grey contours represent the positive field region, and the dotted contours indicate the negative field region. The HCS is represented by the thick solid black line. The pink circle indicates the location of the CME associated with event 7. The dashed and solid pink lines show the CME axis orientations before and after rotation, respectively.}.
\label{f5.6}
\end{figure}

We derived the r.m.s error between observed and predicted MC profiles to estimate the prediction quality. The values of the r.m.s error (see Column 7 of Table \ref{t5.4}) suggest that for most of the cases, the predicted magnetic field magnitude and vector time series agree well with in-situ observations. We note that $\Delta^m_{rms}$ for events 2 and 7 is greater than $\text{twice}$ (average ($\bar{\Delta}^m_{rms}$) $\pm$ standard deviation ($\sigma_{\Delta^m_{rms}}$)) the $\Delta^m_{rms}$ values associated with other events. Although event 2 has a similar FR type in the near-Sun and near-Earth region ($type_{ns} = type_{ne} =$ WNE), a significant asymmetry exists in its magnetic field strength between inbound (while the spacecraft propagates towards the MC centre) and outbound (while the spacecraft propagates away from the MC centre) paths, which might enhance the value of $\Delta^m_{rms}$. The asymmetry does not only occur because of FR expansion or ageing effects \citep{2008SoPh..250..347D, 2018A&A...619A.139D}. Most of the MC field strength asymmetry is instead due to the non-circular cross section of FRs \citep{2009A&A...507..969D}. \citet{janvier2019generic} and \citet{lanabere2020magnetic} quantified the FR asymmetry $C_B$ as $C_B=\frac{\int_{MC_{end}}^{MC_{start}}\frac{t-t_c}{MC_{end}-MC_{start}}B(t) dt}{\int_{MC_{end}}^{MC_{start}} B(t) dt}$, where $B(t)$ is the magnetic field strength and $t_c=(MC_{start}+MC_{end})/2$ represents the central time. Therefore, |$C_B$| increases with magnetic field asymmetry, and a large asymmetry is marked $|C_B| > 0.1$ \citep{lanabere2020magnetic}. We obtain $C_B= -0.12$ in case of event 2, which is greater than $\bar{|C_B|}\pm \delta{|C_B|} = 0.04\pm 0.04$ derived for other events. Here $\bar{|C_B|}$ and $\delta{|C_B|}$ indicate the mean and standard deviation of $C_B$ values, respectively. This implies that a circular cross-section model is inappropriate for estimating its magnetic profile. In Figure \ref{f5.6}(a) we show the in-situ asymmetric magnetic field intensity $B$ of the event 2 MC. The interval within the vertical lines represents the MC interval.
The FR associated with event 7 rotates significantly while propagating from the Sun to Earth and results in a comparatively high value of $\Delta^m_{rms}$. Due to rotation, the FR type changes from the near-Sun to near-Earth regions for this event. By comparing the $\theta^{mva}_{MC}$ and $\theta^{m}_{MC}$ of this event, we find that the FR changes its type from high inclination (the central axis is more or less perpendicular to the ecliptic plane) to low inclination (the central axis is more or less parallel to the ecliptic plane) while propagating in the interplanetary medium. Based on statistical evidence, \citet{yurchyshyn2008relationship, yurchyshyn2009rotation} and \citet{2013SoPh..284..203I} suggested that MCs rotate towards the heliospheric current sheet \citep[HCS;][]{smith2001heliospheric} so that they stay aligned with the local HCS. We considered the Wilcox solar observatory coronal field map calculated from synoptic photospheric magnetogram with a potential field model \citep{hoeksema1983structure, hoeksema1984structure} during Carrington rotation (CR) 2134 when the eruption associated with event 7 occurred. Using $\theta_{HG}$ and $\phi_{HG}$ , we inferred the CME locations on the coronal field map. We observe that in order to stay aligned with the HCS, the associated CME axis underwent significant rotation ($\sim56 ^\circ$) and became more or less parallel to the ecliptic plane by the time it reached at 2.5 $R_\odot$. For context, we show in Figure \ref{f5.6}(b) a coronal map during CR 2134 obtained from the Wilcox Solar Observatory Source Surface Synoptic Charts \footnote{\url{http://wso.stanford.edu/synsourcel.html}}. The solid grey contours represent the positive field region, and the dotted contours indicate the negative field region. The solid thick black line represents the location of the HCS. The pink circle indicates the CME location, and the dotted and solid pink lines show the orientation of the CME axis before and after rotation, respectively.     
\par
  
\citet{sarkar2020observationally} noted that the $B_{x}$ component is more sensitive to small variations ($\pm 10 ^\circ$) in the CME propagation direction and tilt than the $B_{y}$ and $B_{z}$ components. They found that within the propagation direction uncertainties, the $B_x$ component of MCs may have both positive and negative components. In our study we observe that the uncertainty in the propagation direction of the CME and the tilt leads to a significant variation in the predicted $B_{x}$ profiles of MCs associated with events 1, 3, 4, 6, and 10 (see the dotted blue lines in the third panel of Figure \ref{f5.5}(a), (c), (d), (f), and (j), which have both positive and negative values). 
\par
To obtain the deprojected geometrical parameters and kinematics of CMEs, simultaneous observations from different vantage points in space are necessary \citep{bosman2012three}. \citet{bosman20163} demonstrated that to resolve a CME well globally (3D) from 2D plane-of-sky images obtained using coronagraphs on board of spacecraft, the angular separation $\zeta$ between the spacecrafts needs to be large, that is, $10 ^\circ < \zeta \le 90 ^\circ$. If $\zeta$ is in between $0 ^\circ - 10 ^\circ$, the 2D plane-of-sky images obtained from two coronagraphs on board of two separate spacecraft become nearly congruent and the derivation of deprojected CME parameters becomes nearly impossible, whereas a value of $\zeta = 90 ^\circ$ provides the best condition to resolve a CME in 3D \citep{2009SoPh..256..111T}. We here determined the CME properties by observing each CME simultaneously in three coronagraphs on board of three spacecraft that viewed CMEs from three different angles. However, in the absence of STEREO B data, we can obtained 3D parameters of the CMEs using two coronagraphs viewing a CME from two different angles with a separation of more than $10 ^\circ$ . \citet{bosman2012three} observed CMEs from two different viewing angles using the STEREO A and B spacecraft and derived their 3D properties by fitting the GCS model. \citet{chen2019characteristics} and \citet{palmerio2021predicting} used the STEREO A and LASCO coronagraphs to obtain 3D CME properties using the GCS model fit. The out-of-ecliptic observations of Metis, the multi-wavelength coronagraph for the Solar Orbiter mission, and potential L5 and L4 solar missions are expected to have significant contributions in enhancing the precision of CME parametrisation. 
\par

The framework we presented to estimate the magnetic field time evolution of the near-Earth crossing of MCs, their arrival time, and passage duration appears very promising. As discussed in the Introduction, the capability of reliably estimating the time series of $B_z$ is crucial for space weather forecasting. However, this approach is not expected to perform well in some cases such as strongly interacting CMEs, in which CMEs interact significantly with other CMEs and extraneous magnetic transients, for instance when their propagation is influenced by a fast stream originating from nearby coronal holes, when their configuration is influenced by the heliospheric current sheet and the fast solar wind stream, when their cross-sections differ strongly from a circular shape. As we constrained the MCP model inputs by near-Sun observations alone, the model cannot capture any possible influence of CME FR distortions occurring in the interplanetary medium. The outcomes from probing CMEs using different spacecraft (e.g. the Parker solar probe, Solar Orbiter, BepiColombo, MESSENGER, and VEX) at different heliocentric distances smaller than 1 AU can be used to tune the inputs to the model to enhance the model performance. Inputs from MHD models of interacting magnetic structures in the inner heliosphere may provide useful insights in these contexts. 
These added refinements would decrease the lead time in forecasting CME magnetic field at 1 AU, however. The recent studies by \citet{2022ApJ...924L...6M} and \citet{2021A&A...656L...6O} listed and analysed a few events that were consecutively observed using lineups by spacecraft in the interplanetary medium before the events reached 1 AU. Such observational inputs will undoubtedly help improve the MCP model performance at 1 AU. 

\section{Conclusions} \label{sec5.5}
We developed a scheme to predict the time series of magnetic field vectors of CME-associated magnetic clouds during their near-Earth passage and forecast their arrival time, speed, and duration of passage. The CESSI-MCP model is completely constrained by solar disk and near-Sun observations, is computationally fast, and provides a long time window for predictions; therefore, this approach can be easily transitioned to operational forecasting. The ability to perform all these tasks at high fidelity, including predicting the passage duration of MCs, is significant from the space weather perspective. Not only will the CESSI-MCP modelling framework benefit mitigation strategies for space weather, our work will also provide context for and complement currently ongoing missions (ACE, WIND, DSCOVR, PSP, and Solar Orbiter) and upcomings missions such as the Aditya-L1 mission. This is India's first mission to observe the Sun and characterise the near-Earth space environment from the first Lagrange point L1.

The enhanced functional utility of our method is due to a combination of factors, including a realistic constraint of the CME flux rope using solar observations and allowing the expansion of its cross-section. Our work emphasises the importance of near-Sun observations, multi-vantage point observations, in-situ observations, and coupling the pre-established models and techniques to derive realistic intrinsic parameters of CMEs from the Sun to near-Earth space in order to facilitate space weather assessment and forecasts.

\begin{acknowledgements}
The development of the CESSI magnetic cloud prediction (CESSI-MCP) model was performed at the Center of Excellence in Space Sciences India (CESSI) at the Indian Institute of Science Education and Research, Kolkata. S.P. and E.K. acknowledge support from the European Research Council (ERC) under the European Union's Horizon 2020 Research and Innovation Program Project SolMAG 724391 and the frame work for the Finnish Centre of Excellence in Research of Sustainable Space (FORESAIL; Academy of Finland grant numbers 312390). The authors acknowledge the use of data from the STEREO, SDO, SOHO and ACE instruments. The PhD research of S.P. was supported by the Ministry of Education, Government of India.
\end{acknowledgements}
\bibliographystyle{aa}
\bibliography{aa}

\begin{thebibliography}{133}
\expandafter\ifx\csname natexlab\endcsname\relax\def\natexlab#1{#1}\fi

\bibitem[{{Berger}(2005)}]{2005HiA....13...85B}
{Berger}, M.~A. 2005, Highlights of Astronomy, 13, 85

\bibitem[{{Bhowmik} \& {Nandy}(2018)}]{bhowmik2018}
{Bhowmik}, P. \& {Nandy}, D. 2018, Nature Communications, 9, 5209

\bibitem[{Bosman(2016)}]{bosman20163}
Bosman, E. 2016, PhD thesis, Georg-August-Universit{\"a}t G{\"o}ttingen

\bibitem[{Bosman {et~al.}(2012)Bosman, Bothmer, Nistic{\`o}, Vourlidas, Howard,
  \& Davies}]{bosman2012three}
Bosman, E., Bothmer, V., Nistic{\`o}, G., {et~al.} 2012, Solar Physics, 281,
  167

\bibitem[{{Bosman} {et~al.}(2012){Bosman}, {Bothmer}, {Nistic{\`o}},
  {Vourlidas}, {Howard}, \& {Davies}}]{2012SoPh..281..167B}
{Bosman}, E., {Bothmer}, V., {Nistic{\`o}}, G., {et~al.} 2012, Solar Physics,
  281, 167

\bibitem[{Bothmer \& Schwenn(1998)}]{angeo-16-1-1998}
Bothmer, V. \& Schwenn, R. 1998, Annales Geophysicae, 16, 1

\bibitem[{{Brueckner} {et~al.}(1995){Brueckner}, {Howard}, {Koomen},
  {Korendyke}, {Michels}, {Moses}, {Socker}, {Dere}, {Lamy}, {Llebaria},
  {Bout}, {Schwenn}, {Simnett}, {Bedford}, \& {Eyles}}]{1995SoPh..162..357B}
{Brueckner}, G.~E., {Howard}, R.~A., {Koomen}, M.~J., {et~al.} 1995, Solar
  Physics, 162, 357

\bibitem[{Burkepile {et~al.}(2004)Burkepile, Hundhausen, Stanger, St.Cyr, \&
  Seiden}]{JGRA:JGRA17179}
Burkepile, J.~T., Hundhausen, A.~J., Stanger, A.~L., St.Cyr, O.~C., \& Seiden,
  J.~A. 2004, Journal of Geophysical Research (Space Physics), 109

\bibitem[{Burlaga(1988)}]{burlaga1988magnetic}
Burlaga, L. 1988, Journal of Geophysical Research (Space Physics), 93, 7217

\bibitem[{Burlaga(1991)}]{burlaga1991intermittent}
Burlaga, L. 1991, Journal of Geophysical Research (Space Physics), 96, 5847

\bibitem[{{Burlaga} {et~al.}(1981){Burlaga}, {Sittler}, {Mariani}, \&
  {Schwenn}}]{1981JGR....86.6673B}
{Burlaga}, L., {Sittler}, E., {Mariani}, F., \& {Schwenn}, R. 1981, Journal of
  Geophysical Research (Space Physics), 86, 6673

\bibitem[{Burlaga {et~al.}(1981)Burlaga, Sittler, Mariani, \&
  Schwenn}]{1981JGR..86..6673B}
Burlaga, L., Sittler, E., Mariani, F., \& Schwenn, R. 1981, Journal of
  Geophysical Research (Space Physics), 86, 6673

\bibitem[{{\v{C}}alogovi{\'c} {et~al.}(2021){\v{C}}alogovi{\'c}, Dumbovi{\'c},
  Sudar, Vr{\v{s}}nak, Martini{\'c}, Temmer, \&
  Veronig}]{vcalogovic2021probabilistic}
{\v{C}}alogovi{\'c}, J., Dumbovi{\'c}, M., Sudar, D., {et~al.} 2021, Solar
  Physics, 296, 1

\bibitem[{Chen {et~al.}(2019)Chen, Liu, Wang, Zhao, Hu, \&
  Zhu}]{chen2019characteristics}
Chen, C., Liu, Y.~D., Wang, R., {et~al.} 2019, The Astrophysical Journal, 884,
  90

\bibitem[{Cook {et~al.}(2009)Cook, Mackay, \& Nandy}]{cook2009solar}
Cook, G., Mackay, D., \& Nandy, D. 2009, Astrophys. J., 704, 1021

\bibitem[{D{\'e}moulin \& Dasso(2009)}]{demoulin2009causes}
D{\'e}moulin, P. \& Dasso, S. 2009, Astronomy \& Astrophysics, 498, 551

\bibitem[{{D{\'e}moulin} \& {Dasso}(2009)}]{2009A&A...507..969D}
{D{\'e}moulin}, P. \& {Dasso}, S. 2009, \aap, 507, 969

\bibitem[{{D{\'e}moulin} {et~al.}(2018){D{\'e}moulin}, {Dasso}, \&
  {Janvier}}]{2018A&A...619A.139D}
{D{\'e}moulin}, P., {Dasso}, S., \& {Janvier}, M. 2018, \aap, 619, A139

\bibitem[{D{\'e}moulin {et~al.}(2008)D{\'e}moulin, Nakwacki, Dasso, \&
  Mandrini}]{demoulin2008expected}
D{\'e}moulin, P., Nakwacki, M.~S., Dasso, S., \& Mandrini, C.~H. 2008, Solar
  Physics, 250, 347

\bibitem[{{D{\'e}moulin} {et~al.}(2008){D{\'e}moulin}, {Nakwacki}, {Dasso}, \&
  {Mandrini}}]{2008SoPh..250..347D}
{D{\'e}moulin}, P., {Nakwacki}, M.~S., {Dasso}, S., \& {Mandrini}, C.~H. 2008,
  \solphys, 250, 347

\bibitem[{D{\'e}moulin {et~al.}(1996)D{\'e}moulin, Priest, \&
  Lonie}]{demoulin1996three}
D{\'e}moulin, P., Priest, E., \& Lonie, D. 1996, Journal of Geophysical
  Research (Space Physics), 101, 7631

\bibitem[{Dumbovi{\'c} {et~al.}(2018)Dumbovi{\'c}, {\v{C}}alogovi{\'c},
  Vr{\v{s}}nak, Temmer, Mays, Veronig, \& Piantschitsch}]{dumbovic2018drag}
Dumbovi{\'c}, M., {\v{C}}alogovi{\'c}, J., Vr{\v{s}}nak, B., {et~al.} 2018, The
  Astrophysical Journal, 854, 180

\bibitem[{{Dumbovi{\'c}} {et~al.}(2018){Dumbovi{\'c}}, {{\v C}alogovi{\'c}},
  {Vr{\v s}nak}, {Temmer}, {Mays}, {Veronig}, \&
  {Piantschitsch}}]{2018ApJ...854..180D}
{Dumbovi{\'c}}, M., {{\v C}alogovi{\'c}}, J., {Vr{\v s}nak}, B., {et~al.} 2018,
  The Astrophysical Journal, 854, 180

\bibitem[{Fuentes {et~al.}(2000)Fuentes, D{\'e}moulin, Mandrini, \& van
  Driel-Gesztelyi}]{fuentes2000counterkink}
Fuentes, M.~L., D{\'e}moulin, P., Mandrini, C.~H., \& van Driel-Gesztelyi, L.
  2000, The Astrophysical Journal, 544, 540

\bibitem[{{Goldstein}(1983)}]{1983NASCP.2280.731G}
{Goldstein}, H. 1983, in NASA Conference Publication, Vol. 228, NASA Conference
  Publication

\bibitem[{Gonzalez {et~al.}(1999)Gonzalez, Tsurutani, \&
  De~Gonzalez}]{gonzalez1999interplanetary}
Gonzalez, W.~D., Tsurutani, B.~T., \& De~Gonzalez, A. L.~C. 1999, Space Science
  Reviews, 88, 529

\bibitem[{Gopalswamy {et~al.}(2017)Gopalswamy, Akiyama, Yashiro, \&
  Xie}]{gopalswamy2017new}
Gopalswamy, N., Akiyama, S., Yashiro, S., \& Xie, H. 2017, Proceedings of the
  International Astronomical Union, 13, 258

\bibitem[{Gopalswamy {et~al.}(2017b)Gopalswamy, Akiyama, Yashiro, \&
  Xie}]{GOPALSWAMY2017b}
Gopalswamy, N., Akiyama, S., Yashiro, S., \& Xie, H. 2017b, Journal of
  Atmospheric and Solar-Terrestrial Physics

\bibitem[{{Gopalswamy} {et~al.}({2017a}){Gopalswamy}, {Yashiro}, {Akiyama}, \&
  {Xie}}]{2017aSoPh..292...65G}
{Gopalswamy}, N., {Yashiro}, S., {Akiyama}, S., \& {Xie}, H. {2017a}, Solar
  Physics, 292, 65

\bibitem[{Hidalgo(2003)}]{hidalgo2003study}
Hidalgo, M. 2003, Journal of Geophysical Research (Space Physics), 108

\bibitem[{Hoeksema(1984)}]{hoeksema1984structure}
Hoeksema, J.~T. 1984, Structure and Evoluton of the Large Scale Solar and
  Heliospheric Magnetic Fields., Tech. rep., STANFORD UNIV CA CENTER FOR SPACE
  SCIENCE AND ASTROPHYSICS

\bibitem[{Hoeksema {et~al.}(1983)Hoeksema, Wilcox, \&
  Scherrer}]{hoeksema1983structure}
Hoeksema, J.~T., Wilcox, J.~M., \& Scherrer, P.~H. 1983, Journal of Geophysical
  Research: Space Physics, 88, 9910

\bibitem[{{Howard} {et~al.}(2008){Howard}, {Moses}, {Vourlidas}, {Newmark},
  {Socker}, {Plunkett}, {Korendyke}, {Cook}, {Hurley}, {Davila}, {Thompson},
  {St Cyr}, {Mentzell}, {Mehalick}, {Lemen}, {Wuelser}, {Duncan}, {Tarbell},
  {Wolfson}, {Moore}, {Harrison}, {Waltham}, {Lang}, {Davis}, {Eyles},
  {Mapson-Menard}, {Simnett}, {Halain}, {Defise}, {Mazy}, {Rochus}, {Mercier},
  {Ravet}, {Delmotte}, {Auchere}, {Delaboudiniere}, {Bothmer}, {Deutsch},
  {Wang}, {Rich}, {Cooper}, {Stephens}, {Maahs}, {Baugh}, {McMullin}, \&
  {Carter}}]{2008SSRv..136...67H}
{Howard}, R.~A., {Moses}, J.~D., {Vourlidas}, A., {et~al.} 2008, Space Science
  Reviews, 136, 67

\bibitem[{Howard {et~al.}(2008)Howard, Nandy, \& Koepke}]{howard2008kinematic}
Howard, T., Nandy, D., \& Koepke, A. 2008, Journal of Geophysical Research
  (Space Physics), 113

\bibitem[{{Hu} {et~al.}(2014){Hu}, {Qiu}, {Dasgupta}, {Khare}, \&
  {Webb}}]{2014ApJ...793...53H}
{Hu}, Q., {Qiu}, J., {Dasgupta}, B., {Khare}, A., \& {Webb}, G.~M. 2014, The
  Astrophysical Journal, 793, 53

\bibitem[{{Isavnin} {et~al.}(2013){Isavnin}, {Vourlidas}, \&
  {Kilpua}}]{2013SoPh..284..203I}
{Isavnin}, A., {Vourlidas}, A., \& {Kilpua}, E.~K.~J. 2013, \solphys, 284, 203

\bibitem[{{Isavnin} {et~al.}(2014){Isavnin}, {Vourlidas}, \&
  {Kilpua}}]{2014SoPh..289.2141I}
{Isavnin}, A., {Vourlidas}, A., \& {Kilpua}, E.~K.~J. 2014, \solphys, 289, 2141

\bibitem[{Janvier {et~al.}(2019)Janvier, Winslow, Good, Bonhomme, D{\'e}moulin,
  Dasso, M{\"o}stl, Lugaz, Amerstorfer, Soubri{\'e},
  {et~al.}}]{janvier2019generic}
Janvier, M., Winslow, R.~M., Good, S., {et~al.} 2019, Journal of Geophysical
  Research: Space Physics, 124, 812

\bibitem[{Kay \& Gopalswamy(2017)}]{kay2017using}
Kay, C. \& Gopalswamy, N. 2017, Journal of Geophysical Research (Space
  Physics), 122, 11

\bibitem[{{Kay} \& {Gopalswamy}(2018)}]{2018JGRA..123.7220K}
{Kay}, C. \& {Gopalswamy}, N. 2018, Journal of Geophysical Research (Space
  Physics), 123, 7220

\bibitem[{Kay {et~al.}(2017)Kay, Gopalswamy, Reinard, \&
  Opher}]{kay2017predicting}
Kay, C., Gopalswamy, N., Reinard, A., \& Opher, M. 2017, The Astrophysical
  Journal, 835, 117

\bibitem[{Kay {et~al.}(2022)Kay, Mays, \&
  Collado-Vega}]{https://doi.org/10.1029/2021SW002914}
Kay, C., Mays, M.~L., \& Collado-Vega, Y.~M. 2022, Space Weather, 20,
  e2021SW002914, e2021SW002914 2021SW002914

\bibitem[{Kay \& Opher(2015)}]{kay2015heliocentric}
Kay, C. \& Opher, M. 2015, The Astrophysical Journal Letters, 811, L36

\bibitem[{Kay {et~al.}(2013)Kay, Opher, \& Evans}]{kay2013forecasting}
Kay, C., Opher, M., \& Evans, R.~M. 2013, The Astrophysical Journal, 775, 5

\bibitem[{{Kay} {et~al.}(2015){Kay}, {Opher}, \& {Evans}}]{2015ApJ...805..168K}
{Kay}, C., {Opher}, M., \& {Evans}, R.~M. 2015, \apj, 805, 168

\bibitem[{{Kilpua} {et~al.}(2017){Kilpua}, {Koskinen}, \&
  {Pulkkinen}}]{Kilpua2017a}
{Kilpua}, E., {Koskinen}, H. E.~J., \& {Pulkkinen}, T.~I. 2017, Living Reviews
  in Solar Physics, 14, 5

\bibitem[{{Kilpua} {et~al.}(2013){Kilpua}, {Isavnin}, {Vourlidas}, {Koskinen},
  \& {Rodriguez}}]{2013AnGeo..31.1251K}
{Kilpua}, E.~K.~J., {Isavnin}, A., {Vourlidas}, A., {Koskinen}, H.~E.~J., \&
  {Rodriguez}, L. 2013, Annales Geophysicae, 31, 1251

\bibitem[{{Kilpua} {et~al.}(2019){Kilpua}, {Lugaz}, {Mays}, \&
  {Temmer}}]{Kilpua2019}
{Kilpua}, E.~K.~J., {Lugaz}, N., {Mays}, M.~L., \& {Temmer}, M. 2019, Space
  Weather, 17, 498

\bibitem[{Lanabere {et~al.}(2020)Lanabere, Dasso, D{\'e}moulin, Janvier,
  Rodriguez, \& Mas{\'\i}as-Meza}]{lanabere2020magnetic}
Lanabere, V., Dasso, S., D{\'e}moulin, P., {et~al.} 2020, Astronomy \&
  Astrophysics, 635, A85

\bibitem[{{Lepping} {et~al.}(2008){Lepping}, {Wu}, {Berdichevsky}, \&
  {Ferguson}}]{lepping2008estimates}
{Lepping}, R.~P., {Wu}, C.~C., {Berdichevsky}, D.~B., \& {Ferguson}, T. 2008,
  Annales Geophysicae, 26, 1919

\bibitem[{Liu {et~al.}(2014)Liu, Hoeksema, Bobra, Hayashi, Schuck, \&
  Sun}]{liu2014magnetic}
Liu, Y., Hoeksema, J., Bobra, M., {et~al.} 2014, The Astrophysical Journal,
  785, 13

\bibitem[{{Longcope} {et~al.}(2007){Longcope}, {Beveridge}, {Qiu}, {Ravindra},
  {Barnes}, \& {Dasso}}]{2007SoPh..244...45L}
{Longcope}, D., {Beveridge}, C., {Qiu}, J., {et~al.} 2007, Solar Physics, 244,
  45

\bibitem[{Lundquist(1951)}]{PhysRev.83.307}
Lundquist, S. 1951, Phys. Rev., 83, 307

\bibitem[{Luoni {et~al.}(2011)Luoni, D{\'e}moulin, Mandrini, \& van
  Driel-Gesztelyi}]{luoni2011twisted}
Luoni, M.~L., D{\'e}moulin, P., Mandrini, C.~H., \& van Driel-Gesztelyi, L.
  2011, Solar Physics, 270, 45

\bibitem[{Lynch {et~al.}(2009)Lynch, Antiochos, Li, Luhmann, \&
  DeVore}]{lynch2009rotation}
Lynch, B., Antiochos, S., Li, Y., Luhmann, J., \& DeVore, C. 2009, The
  Astrophysical Journal, 697, 1918

\bibitem[{{Manchester} {et~al.}(2017){Manchester}, {Kilpua}, {Liu}, {Lugaz},
  {Riley}, {T{\"o}r{\"o}k}, \& {Vr{\v s}nak}}]{2017SSRv..212.1159M}
{Manchester}, W., {Kilpua}, E.~K.~J., {Liu}, Y.~D., {et~al.} 2017, Space
  Science Reviews, 212, 1159

\bibitem[{{Manchester} {et~al.}(2004){Manchester}, {Gombosi}, {Roussev},
  {Ridley}, {de Zeeuw}, {Sokolov}, {Powell}, \&
  {T{\'o}th}}]{2004JGRA..109.2107M}
{Manchester}, W.~B., {Gombosi}, T.~I., {Roussev}, I., {et~al.} 2004, Journal of
  Geophysical Research (Space Physics), 109, A02107

\bibitem[{Manchester {et~al.}(2014)Manchester, van~der Holst, \&
  Lavraud}]{Manchester_2014}
Manchester, W.~B., van~der Holst, B., \& Lavraud, B. 2014, Plasma Physics and
  Controlled Fusion, 56, 064006

\bibitem[{Mandrini {et~al.}(2005)Mandrini, Pohjolainen, Dasso, Green,
  D{\'e}moulin, van Driel-Gesztelyi, Copperwheat, \&
  Foley}]{mandrini2005interplanetary}
Mandrini, C.~H., Pohjolainen, S., Dasso, S., {et~al.} 2005, Astronomy \&
  Astrophysics, 434, 725

\bibitem[{Martin \& McAllister(1996)}]{martin1996skew}
Martin, S. \& McAllister, A. 1996, in Magnetodynamic phenomena in the solar
  atmosphere (Springer), 497--498

\bibitem[{Martin \& McAllister(1997)}]{martin1997predicting}
Martin, S. \& McAllister, A. 1997, Washington DC American Geophysical Union
  Geophysical Monograph Series, 99, 127

\bibitem[{Martin(2003)}]{martin2003signs}
Martin, S.~F. 2003, Advances in Space Research, 32, 1883

\bibitem[{{Marubashi}(1986)}]{1986AdSpR...6..335M}
{Marubashi}, K. 1986, Advances in Space Research, 6, 335

\bibitem[{Marubashi(1997)}]{marubashi1997interplanetary}
Marubashi, K. 1997, Coronal mass ejections, 99, 147

\bibitem[{Marubashi {et~al.}(2015)Marubashi, Akiyama, Yashiro, Gopalswamy, Cho,
  \& Park}]{2015SoPh..290..1371M}
Marubashi, K., Akiyama, S., Yashiro, S., {et~al.} 2015, Solar Physics, 290,
  1371

\bibitem[{Marubashi \& Lepping(2007)}]{2007AG..25..2453M}
Marubashi, K. \& Lepping, R.~P. 2007, Annales Geophysicae, 25, 2453

\bibitem[{Mays {et~al.}(2015)Mays, Taktakishvili, Pulkkinen, MacNeice,
  Rast{\"a}tter, Odstrcil, Jian, Richardson, LaSota, Zheng,
  {et~al.}}]{mays2015ensemble}
Mays, M., Taktakishvili, A., Pulkkinen, A., {et~al.} 2015, Solar Physics, 290,
  1775

\bibitem[{McAllister {et~al.}(1995)McAllister, Hundhausen, Burkpile, McIntosh,
  \& Hiei}]{mcallister1995declining}
McAllister, A., Hundhausen, A., Burkpile, J., McIntosh, P., \& Hiei, E. 1995,
  in Bulletin of the American Astronomical Society, Vol.~27, 961

\bibitem[{M{\"o}stl {et~al.}(2018)M{\"o}stl, Amerstorfer, Palmerio, Isavnin,
  Farrugia, Lowder, Winslow, Donnerer, Kilpua, \& Boakes}]{mostl2018forward}
M{\"o}stl, C., Amerstorfer, T., Palmerio, E., {et~al.} 2018, Space Weather, 16,
  216

\bibitem[{{M{\"o}stl} {et~al.}(2015){M{\"o}stl}, {Rollett}, {Frahm}, {Liu},
  {Long}, {Colaninno}, {Reiss}, {Temmer}, {Farrugia}, {Posner}, {Dumbovi{\'c}},
  {Janvier}, {D{\'e}moulin}, {Boakes}, {Devos}, {Kraaikamp}, {Mays}, \&
  {Vr{\v{s}}nak}}]{Mostl2015}
{M{\"o}stl}, C., {Rollett}, T., {Frahm}, R.~A., {et~al.} 2015, Nature
  Communications, 6, 7135

\bibitem[{{M{\"o}stl} {et~al.}(2022){M{\"o}stl}, {Weiss}, {Reiss},
  {Amerstorfer}, {Bailey}, {Hinterreiter}, {Bauer}, {Barnes}, {Davies},
  {Harrison}, {Freiherr von Forstner}, {Davies}, {Heyner}, {Horbury}, \&
  {Bale}}]{2022ApJ...924L...6M}
{M{\"o}stl}, C., {Weiss}, A.~J., {Reiss}, M.~A., {et~al.} 2022, \apjl, 924, L6

\bibitem[{Mulligan {et~al.}(1998)Mulligan, Russell, \&
  Luhmann}]{mulligan1998solar}
Mulligan, T., Russell, C., \& Luhmann, J. 1998, Geophysical Research Letters,
  25, 2959

\bibitem[{Nandy(2006)}]{nandy2006magnetic}
Nandy, D. 2006, Journal of Geophysical Research: Space Physics, 111

\bibitem[{{Nandy}(2021)}]{nandy2021a}
{Nandy}, D. 2021, \solphys, 296, 54

\bibitem[{{Nandy} {et~al.}(2021){Nandy}, {Martens}, {Obridko}, {Dash}, \&
  {Georgieva}}]{nandy2021b}
{Nandy}, D., {Martens}, P. C.~H., {Obridko}, V., {Dash}, S., \& {Georgieva}, K.
  2021, Progress in Earth and Planetary Science, 8, 40

\bibitem[{Nieves-Chinchilla {et~al.}(2018)Nieves-Chinchilla, Vourlidas,
  Raymond, Linton, Al-Haddad, Savani, Szabo, \&
  Hidalgo}]{nieves2018understanding}
Nieves-Chinchilla, T., Vourlidas, A., Raymond, J., {et~al.} 2018, Solar
  Physics, 293, 1

\bibitem[{Odstrcil {et~al.}(2004)Odstrcil, Riley, \&
  Zhao}]{odstrcil2004numerical}
Odstrcil, D., Riley, P., \& Zhao, X. 2004, Journal of Geophysical Research
  (Space Physics), 109

\bibitem[{{O'Kane} {et~al.}(2021){O'Kane}, {Green}, {Davies}, {M{\"o}stl},
  {Hinterreiter}, {Freiherr von Forstner}, {Weiss}, {Long}, \&
  {Amerstorfer}}]{2021A&A...656L...6O}
{O'Kane}, J., {Green}, L.~M., {Davies}, E.~E., {et~al.} 2021, \aap, 656, L6

\bibitem[{Osherovich {et~al.}(1993)Osherovich, Farrugia, \&
  Burlaga}]{osherovich1993nonlinear}
Osherovich, V., Farrugia, C., \& Burlaga, L. 1993, Journal of Geophysical
  Research (Space Physics), 98, 13225

\bibitem[{Pal(2021)}]{PAL2021}
Pal, S. 2021, Advances in Space Research

\bibitem[{Pal {et~al.}(2020)Pal, Dash, \& Nandy}]{pal2020flux}
Pal, S., Dash, S., \& Nandy, D. 2020, Geophysical Research Letters, 47,
  e2019GL086372

\bibitem[{{Pal} {et~al.}(2017){Pal}, {Gopalswamy}, {Nandy}, {Akiyama},
  {Yashiro}, {Makela}, \& {Xie}}]{2017ApJ...851..123P}
{Pal}, S., {Gopalswamy}, N., {Nandy}, D., {et~al.} 2017, \apj, 851, 123

\bibitem[{{Pal} {et~al.}(2021){Pal}, {Kilpua}, {Good}, {Pomoell}, \&
  {Price}}]{2021A&A...650A.176P}
{Pal}, S., {Kilpua}, E., {Good}, S., {Pomoell}, J., \& {Price}, D.~J. 2021,
  \aap, 650, A176

\bibitem[{Pal {et~al.}(2022)Pal, Lynch, Good, Palmerio, Asvestari, Pomoell,
  Stevens, \& Kilpua}]{pal2022eruption}
Pal, S., Lynch, B.~J., Good, S.~W., {et~al.} 2022, arXiv preprint
  arXiv:2205.07713

\bibitem[{Pal {et~al.}(2018)Pal, Nandy, Srivastava, Gopalswamy, \&
  Panda}]{pal2018dependence}
Pal, S., Nandy, D., Srivastava, N., Gopalswamy, N., \& Panda, S. 2018, The
  Astrophysical Journal, 865, 4

\bibitem[{{Palmerio} {et~al.}(2021){Palmerio}, {Kay}, {Al-Haddad}, {Lynch},
  {Yu}, {Stevens}, {Pal}, \& {Lee}}]{2021ApJ...920...65P}
{Palmerio}, E., {Kay}, C., {Al-Haddad}, N., {et~al.} 2021, \apj, 920, 65

\bibitem[{Palmerio {et~al.}(2021)Palmerio, Kay, Al-Haddad, Lynch, Yu, Stevens,
  Pal, \& Lee}]{palmerio2021predicting}
Palmerio, E., Kay, C., Al-Haddad, N., {et~al.} 2021, The Astrophysical Journal,
  920, 65

\bibitem[{Palmerio {et~al.}(2017)Palmerio, Kilpua, James, Green, Pomoell,
  Isavnin, \& Valori}]{palmerio2017determining}
Palmerio, E., Kilpua, E.~K., James, A.~W., {et~al.} 2017, Solar Physics, 292,
  39

\bibitem[{Palmerio {et~al.}(2018)Palmerio, Kilpua, M{\"o}stl, Bothmer, James,
  Green, Isavnin, Davies, \& Harrison}]{palmerio2018coronal}
Palmerio, E., Kilpua, E.~K., M{\"o}stl, C., {et~al.} 2018, Space Weather, 16,
  442

\bibitem[{Pevtsov \& Balasubramaniam(2003)}]{pevtsov2003helicity}
Pevtsov, A. \& Balasubramaniam, K. 2003, Advances in Space Research, 32, 1867

\bibitem[{{Pevtsov} {et~al.}(2014){Pevtsov}, {Berger}, {Nindos}, {Norton}, \&
  {van Driel-Gesztelyi}}]{2014SSRv..186..285P}
{Pevtsov}, A.~A., {Berger}, M.~A., {Nindos}, A., {Norton}, A.~A., \& {van
  Driel-Gesztelyi}, L. 2014, \ssr, 186, 285

\bibitem[{{Qiu} {et~al.}(2007){Qiu}, {Hu}, {Howard}, \&
  {Yurchyshyn}}]{2007ApJ...659..758Q}
{Qiu}, J., {Hu}, Q., {Howard}, T.~A., \& {Yurchyshyn}, V.~B. 2007, The
  Astrophysical Journal, 659, 758

\bibitem[{Reiss {et~al.}(2016)Reiss, Temmer, Veronig, Nikolic, Vennerstrom,
  Sch{\"o}ngassner, \& Hofmeister}]{reiss2016verification}
Reiss, M.~A., Temmer, M., Veronig, A.~M., {et~al.} 2016, Space Weather, 14, 495

\bibitem[{Richardson \& Cane(2010)}]{Richardson2010}
Richardson, I.~G. \& Cane, H.~V. 2010, Solar Physics, 264, 189

\bibitem[{{Rodr{\'\i}guez G{\'o}mez} {et~al.}(2020){Rodr{\'\i}guez G{\'o}mez},
  {Podladchikova}, {Veronig}, {Ruzmaikin}, {Feynman}, \&
  {Petrukovich}}]{2020ApJ...899...47R}
{Rodr{\'\i}guez G{\'o}mez}, J.~M., {Podladchikova}, T., {Veronig}, A., {et~al.}
  2020, \apj, 899, 47

\bibitem[{Rotter {et~al.}(2012)Rotter, Veronig, Temmer, \&
  Vr{\v{s}}nak}]{rotter2012relation}
Rotter, T., Veronig, A., Temmer, M., \& Vr{\v{s}}nak, B. 2012, Solar Physics,
  281, 793

\bibitem[{Rotter {et~al.}(2015)Rotter, Veronig, Temmer, \&
  Vr{\v{s}}nak}]{rotter2015real}
Rotter, T., Veronig, A., Temmer, M., \& Vr{\v{s}}nak, B. 2015, Solar Physics,
  290, 1355

\bibitem[{Sarkar {et~al.}(2020)Sarkar, Gopalswamy, \&
  Srivastava}]{sarkar2020observationally}
Sarkar, R., Gopalswamy, N., \& Srivastava, N. 2020, The Astrophysical Journal,
  888, 121

\bibitem[{{Savani} {et~al.}(2015){Savani}, {Vourlidas}, {Szabo}, {Mays},
  {Richardson}, {Thompson}, {Pulkkinen}, {Evans}, \&
  {Nieves-Chinchilla}}]{2015SpWea..13..374S}
{Savani}, N.~P., {Vourlidas}, A., {Szabo}, A., {et~al.} 2015, Space Weather,
  13, 374

\bibitem[{Schrijver {et~al.}(2015)Schrijver, Kauristie, Aylward, Denardini,
  Gibson, Glover, Gopalswamy, Grande, Hapgood, Heynderickx,
  {et~al.}}]{schrijver2015understanding}
Schrijver, C.~J., Kauristie, K., Aylward, A.~D., {et~al.} 2015, Advances in
  Space Research, 55, 2745

\bibitem[{{Scolini} {et~al.}(2021){Scolini}, {Dasso}, {Rodriguez}, {Zhukov}, \&
  {Poedts}}]{2021A&A...649A..69S}
{Scolini}, C., {Dasso}, S., {Rodriguez}, L., {Zhukov}, A.~N., \& {Poedts}, S.
  2021, \aap, 649, A69

\bibitem[{{Shen} {et~al.}(2013){Shen}, {Wang}, {Pan}, {Zhang}, {Ye}, \&
  {Wang}}]{2013JGRA..118.6858S}
{Shen}, C., {Wang}, Y., {Pan}, Z., {et~al.} 2013, Journal of Geophysical
  Research (Space Physics), 118, 6858

\bibitem[{Shen {et~al.}(2014)Shen, Shen, Zhang, Hess, Wang, Feng, Cheng, \&
  Yang}]{shen2014evolution}
Shen, F., Shen, C., Zhang, J., {et~al.} 2014, Journal of Geophysical Research:
  Space Physics, 119, 7128

\bibitem[{Shimazu \& Vandas(2002)}]{shimazu2002self}
Shimazu, H. \& Vandas, M. 2002, Earth, planets and space, 54, 783

\bibitem[{{Sinha} {et~al.}(2022){Sinha}, {Gupta}, {Singh}, {Lekshmi}, {Pal},
  {Nandy}, {Mitra}, {Chatterjee}, {Bhattacharya}, {Chatterjee}, {Srivastava},
  \& {Brandenburg}}]{sinha2022}
{Sinha}, S., {Gupta}, O., {Singh}, V., {et~al.} 2022, arXiv e-prints,
  arXiv:2204.05910

\bibitem[{Smith(2001)}]{smith2001heliospheric}
Smith, E.~J. 2001, Journal of Geophysical Research: Space Physics, 106, 15819

\bibitem[{{Sonnerup} \& {Cahill}(1967)}]{1967JGR....72..171S}
{Sonnerup}, B.~U.~O. \& {Cahill}, L.~J., J. 1967, \jgr, 72, 171

\bibitem[{Subramanian {et~al.}(2014)Subramanian, Arunbabu, Vourlidas, \&
  Mauriya}]{subramanian2014self}
Subramanian, P., Arunbabu, K., Vourlidas, A., \& Mauriya, A. 2014, The
  Astrophysical Journal, 790, 125

\bibitem[{{Thernisien}(2011)}]{2011ApJS..194...33T}
{Thernisien}, A. 2011, The Astrophysical Journals, 194, 33

\bibitem[{{Thernisien} {et~al.}(2009){Thernisien}, {Vourlidas}, \&
  {Howard}}]{2009SoPh..256..111T}
{Thernisien}, A., {Vourlidas}, A., \& {Howard}, R.~A. 2009, Solar Physics, 256,
  111

\bibitem[{{Thernisien} {et~al.}(2006){Thernisien}, {Howard}, \&
  {Vourlidas}}]{2006ApJ...652..763T}
{Thernisien}, A.~F.~R., {Howard}, R.~A., \& {Vourlidas}, A. 2006, The
  Astrophysical Journal, 652, 763

\bibitem[{Tsurutani {et~al.}(1988)Tsurutani, Gonzalez, Tang, Akasofu, \&
  Smith}]{tsurutani1988origin}
Tsurutani, B.~T., Gonzalez, W.~D., Tang, F., Akasofu, S.~I., \& Smith, E.~J.
  1988, Journal of Geophysical Research (Space Physics), 93, 8519

\bibitem[{Vandas {et~al.}(2015)Vandas, Romashets, \&
  Geranios}]{vandas2015modeling}
Vandas, M., Romashets, E., \& Geranios, A. 2015, Astronomy \& Astrophysics,
  583, A78

\bibitem[{Vandas {et~al.}(2006)Vandas, Romashets, Watari, Geranios, Antoniadou,
  \& Zacharopoulou}]{vandas2006comparison}
Vandas, M., Romashets, E., Watari, S., {et~al.} 2006, Advances in Space
  Research, 38, 441

\bibitem[{{Vourlidas} {et~al.}(2017){Vourlidas}, {Balmaceda}, {Stenborg}, \&
  {Dal Lago}}]{Vourlidas2017}
{Vourlidas}, A., {Balmaceda}, L.~A., {Stenborg}, G., \& {Dal Lago}, A. 2017,
  \apj, 838, 141

\bibitem[{Vourlidas {et~al.}(2011)Vourlidas, Colaninno, Nieves-Chinchilla, \&
  Stenborg}]{vourlidas2011first}
Vourlidas, A., Colaninno, R., Nieves-Chinchilla, T., \& Stenborg, G. 2011, The
  Astrophysical Journal Letters, 733, L23

\bibitem[{Vr{\v{s}}nak {et~al.}(2019)Vr{\v{s}}nak, Amerstorfer, Dumbovi{\'c},
  Leitner, Veronig, Temmer, M{\"o}stl, Amerstorfer, Farrugia, \&
  Galvin}]{vrvsnak2019heliospheric}
Vr{\v{s}}nak, B., Amerstorfer, T., Dumbovi{\'c}, M., {et~al.} 2019, The
  Astrophysical Journal, 877, 77

\bibitem[{Vr{\v{s}}nak {et~al.}(2007)Vr{\v{s}}nak, Temmer, \&
  Veronig}]{vrvsnak2007coronal}
Vr{\v{s}}nak, B., Temmer, M., \& Veronig, A.~M. 2007, Solar Physics, 240, 315

\bibitem[{Vr{\v{s}}nak {et~al.}(2014)Vr{\v{s}}nak, Temmer, {\v{Z}}ic,
  Taktakishvili, Dumbovi{\'c}, M{\"o}stl, Veronig, Mays, \&
  Odstr{\v{c}}il}]{vrvsnak2014heliospheric}
Vr{\v{s}}nak, B., Temmer, M., {\v{Z}}ic, T., {et~al.} 2014, The Astrophysical
  Journal Supplement Series, 213, 21

\bibitem[{Vr{\v{s}}nak \& {\v{Z}}ic(2007)}]{vrvsnak2007transit}
Vr{\v{s}}nak, B. \& {\v{Z}}ic, T. 2007, Astronomy \& Astrophysics, 472, 937

\bibitem[{Vr{\v{s}}nak {et~al.}(2010)Vr{\v{s}}nak, {\v{Z}}ic, Falkenberg,
  M{\"o}stl, Vennerstrom, \& Vrbanec}]{vrvsnak2010role}
Vr{\v{s}}nak, B., {\v{Z}}ic, T., Falkenberg, T.~V., {et~al.} 2010, Astronomy \&
  Astrophysics, 512, A43

\bibitem[{Vr{\v{s}}nak {et~al.}(2013)Vr{\v{s}}nak, {\v{Z}}ic, Vrbanec, Temmer,
  Rollett, M{\"o}stl, Veronig, {\v{C}}alogovi{\'c}, Dumbovi{\'c}, Luli{\'c},
  {et~al.}}]{vrvsnak2013propagation}
Vr{\v{s}}nak, B., {\v{Z}}ic, T., Vrbanec, D., {et~al.} 2013, Solar physics,
  285, 295

\bibitem[{{Vr{\v s}nak} {et~al.}(2013){Vr{\v s}nak}, {{\v Z}ic}, {Vrbanec},
  {Temmer}, {Rollett}, {M{\"o}stl}, {Veronig}, {{\v C}alogovi{\'c}},
  {Dumbovi{\'c}}, {Luli{\'c}}, {Moon}, \& {Shanmugaraju}}]{2013SoPh..285..295V}
{Vr{\v s}nak}, B., {{\v Z}ic}, T., {Vrbanec}, D., {et~al.} 2013, Solar Physics,
  285, 295

\bibitem[{{Vr{\v{s}}nak} {et~al.}(2010){Vr{\v{s}}nak}, {{\v{Z}}ic},
  {Falkenberg}, {M{\"o}stl}, {Vennerstrom}, \& {Vrbanec}}]{2010A&A...512A..43V}
{Vr{\v{s}}nak}, B., {{\v{Z}}ic}, T., {Falkenberg}, T.~V., {et~al.} 2010, \aap,
  512, A43

\bibitem[{Wang {et~al.}(2004)Wang, Shen, Wang, \& Ye}]{wang2004deflection}
Wang, Y., Shen, C., Wang, S., \& Ye, P. 2004, Solar Physics, 222, 329

\bibitem[{{Wang} {et~al.}(2004){Wang}, {Shen}, {Wang}, \&
  {Ye}}]{2004SoPh..222..329W}
{Wang}, Y., {Shen}, C., {Wang}, S., \& {Ye}, P. 2004, \solphys, 222, 329

\bibitem[{{Wang} {et~al.}(2014){Wang}, {Wang}, {Shen}, {Shen}, \&
  {Lugaz}}]{2014JGRA..119.5117W}
{Wang}, Y., {Wang}, B., {Shen}, C., {Shen}, F., \& {Lugaz}, N. 2014, Journal of
  Geophysical Research (Space Physics), 119, 5117

\bibitem[{{Webb} \& {Howard}(2012)}]{Webb2012}
{Webb}, D.~F. \& {Howard}, T.~A. 2012, Living Reviews in Solar Physics, 9, 3

\bibitem[{Yeates {et~al.}(2010)Yeates, Attrill, Nandy, Mackay, Martens, \& van
  Ballegooijen}]{yeates2010comparison}
Yeates, A., Attrill, G., Nandy, D., {et~al.} 2010, The Astrophysical Journal,
  709, 1238

\bibitem[{Yurchyshyn(2008)}]{yurchyshyn2008relationship}
Yurchyshyn, V. 2008, The Astrophysical Journal Letters, 675, L49

\bibitem[{Yurchyshyn {et~al.}(2009)Yurchyshyn, Abramenko, \&
  Tripathi}]{yurchyshyn2009rotation}
Yurchyshyn, V., Abramenko, V., \& Tripathi, D. 2009, The Astrophysical Journal,
  705, 426

\bibitem[{{Zhang} {et~al.}(2007){Zhang}, {Richardson}, {Webb}, {Gopalswamy},
  {Huttunen}, {Kasper}, {Nitta}, {Poomvises}, {Thompson}, {Wu}, {Yashiro}, \&
  {Zhukov}}]{2007JGRA..11212103Z}
{Zhang}, J., {Richardson}, I.~G., {Webb}, D.~F., {et~al.} 2007, Journal of
  Geophysical Research (Space Physics), 112, A12103

\bibitem[{{\v{Z}}ic {et~al.}(2015){\v{Z}}ic, Vr{\v{s}}nak, \&
  Temmer}]{vzic2015heliospheric}
{\v{Z}}ic, T., Vr{\v{s}}nak, B., \& Temmer, M. 2015, The Astrophysical Journal
  Supplement Series, 218, 32

\end{thebibliography}
\newpage


\end{document}